\documentclass[traditabstract]{aa}
\pdfoutput=1
\usepackage{graphicx}
\usepackage{amsmath,amsfonts,amssymb,tabu}
\usepackage{txfonts}
\usepackage[breaklinks,colorlinks,citecolor=blue,pdfa=true,backref=page]{hyperref}
\usepackage{color}
\usepackage{fixltx2e}
\usepackage{natbib,twoopt}
\usepackage{url}
\usepackage{multirow}
\usepackage{epsf}
\usepackage{epsfig}
\usepackage{longtable}
\usepackage{float}
\usepackage{subfig}
\usepackage{caption}
\usepackage[dvipsnames]{xcolor}
\usepackage{upgreek}
\usepackage{caption}

\definecolor{mypink1}{RGB}{219, 48, 122}

\usepackage{ifthen}
\usepackage[T1]{fontenc}
\usepackage{lmodern}
\usepackage{ifxetex,ifluatex}
\usepackage{latexsym}
\usepackage{pdfpages}
\usepackage{dblfloatfix}
\usepackage{morefloats}
\usepackage{caption}
\usepackage{lscape}
\usepackage{mathtools}

\bibpunct{(}{)}{;}{a}{}{,} %% natbib format for A&A and ApJ
\makeatletter
\newcommandtwoopt{\citeads}[3][][]{\href{http://adsabs.harvard.edu/abs/#3}%
{\def\hyper@linkstart##1##2{}%
\let\hyper@linkend\@empty\citealp[#1][#2]{#3}}}
\newcommandtwoopt{\citepads}[3][][]{\href{http://adsabs.harvard.edu/abs/#3}%
{\def\hyper@linkstart##1##2{}%
\let\hyper@linkend\@empty\citep[#1][#2]{#3}}}
\newcommandtwoopt{\citetads}[3][][]{\href{http://adsabs.harvard.edu/abs/#3}%
{\def\hyper@linkstart##1##2{}%
\let\hyper@linkend\@empty\citet[#1][#2]{#3}}}
\newcommandtwoopt{\citeyearads}[3][][]%
{\href{http://adsabs.harvard.edu/abs/#3}
{\def\hyper@linkstart##1##2{}%
\let\hyper@linkend\@empty\citeyear[#1][#2]{#3}}}
\makeatother

\usepackage{graphicx}
\def \deg         {\text{$^{\circ}$}}

\begin{document} 

\title{Barbell-shaped giant radio galaxy with $\sim$\,100 kpc kink in the jet}
\titlerunning{The Barbell GRG}

\author {Pratik Dabhade\inst{1}\thanks{E-mail: pratik.dabhade@obspm.fr, pratikdabhade13@gmail.com}
\and Timothy W. Shimwell\inst{2,3}
\and Joydeep Bagchi\inst{4,5}
\and D. J. Saikia\inst{4,6}
\and Fran\c{c}oise Combes\inst{1}\\
\and Madhuri Gaikwad\inst{7}
\and H. J. A. R\"{o}ttgering\inst{3}
\and Abhisek Mohapatra\inst{4}
\and C. H. Ishwara-Chandra\inst{8}\\
\and Huib T. Intema\inst{3}
\and Somak Raychaudhury\inst{4,9}
}

% Joydeep Bagchi, Timothy Shimwell, D.J.Saikia, Ishwara-Chandra, Huub Rottgering, Madhuri Gaikwad , Huib Intema 

\institute{
$^{1}$Observatoire de Paris, LERMA, Coll\`ege de France, CNRS, PSL University, Sorbonne University, 75014, Paris, France\\
$^{2}$ASTRON, the Netherlands Institute for Radio Astronomy, Postbus 2, NL-7990 AA Dwingeloo, the Netherlands \\
$^{3}$Leiden Observatory, Leiden University, P.O. Box 9513, NL-2300 RA, Leiden, The Netherlands\\
$^{4}$Inter-University Centre for Astronomy and Astrophysics (IUCAA), Pune 411007, India\\ 
$^{5}$Department of Physics and Electronics, CHRIST (Deemed to be University), Bengaluru-560029, India\\
$^{6}$Department of Physics, Tezpur University, Tezpur 784028, India\\
$^{7}$ Max-Planck-Institut f\"ur Radioastronomie, Auf dem Hugel 69, 53121 Bonn, Germany \\
$^{8}$National Centre for Radio Astrophysics, TIFR, Post Bag 3, Ganeshkhind, Pune - 411007, India \\
$^{9}$School of Physics and Astronomy, University of Birmingham, Birmingham B15~2TT, UK}

\authorrunning{Dabhade et al.}
% \abstract{}{}{}{}{} 
% 5 {} token are mandatory
 \date{\today} 
 
  \abstract{We present  for the first time a study of  peculiar giant radio galaxy (GRG)  J223301+131502 using deep multi-frequency radio observations from GMRT (323, 612, and 1300 MHz) and LOFAR (144 MHz) along with optical spectroscopic observations with the WHT~4.2m optical telescope. Our observations have firmly established its redshift of 0.09956 and unveiled its exceptional jet structure extending more than $\sim$\,200\,kpc leading to a peculiar kink structure of $\sim$\,100\,kpc. We measure the overall size of this GRG to be $\sim$\,1.83\,Mpc; it exhibits lobes without any prominent hotspots and closely resembles a barbell.
  Our deep low-frequency radio maps clearly reveal the steep-spectrum diffuse emission from the lobes of the GRG.
  The magnetic field strength of $\sim$\,5\,$\muup$G and spectral ages between about 110 to 200 mega years for the radio lobes were estimated using radio data from LOFAR 144 MHz observations and GMRT 323 and 612 MHz observations. 
  We discuss the possible causes leading to the formation of the observed kink feature for the GRG, which include precession of the jet axis, development of instabilities and magnetic reconnection. Despite its enormous size, the Barbell GRG is found to be residing in a low-mass (M$_{200} \sim 10^{14}$ $\rm M_{\odot}$) galaxy cluster. This GRG with two-sided large-scale jets with a kink and diffuse outer lobes residing in a cluster environment, provides an opportunity to explore the structure and growth of GRGs in different environments.}


\maketitle

\section{Introduction} \label{sec:intro}
Radio-loud active galactic nuclei (AGNs) are classified  based on their observed properties. Two of these are radio galaxies (RGs) and radio quasars (RQs).
RGs and RQs often exhibit relativistic radio jets connecting the nucleus to the outer extended emission.

Extended RGs and RQs with edge-brightened lobes are categorised as Fanaroff-Riley type II (FRII) \citep{FR74} and those with centre-brightened features with bright jets and no hotspots are called  Fanaroff-Riley type I (FRI) sources. The launching and propagation of jets are explained by the standard beam model \citep{longair73,Scheuer74,Blanford_Rees74}, where collimated jets resulting from outflows originating from supermassive black holes give rise to the extended radio structure. The dichotomy between the FRI and FRII  sources is thought to be rooted in the AGN accretion mode, in  connection with  jet power, gas entrainment, and possibly with the  properties of the host galaxies \citep{bh12rgs,mingo14,hardcastlenat,Mingo22}. 

The properties of jets in radio-loud AGN (RLAGN), which include both RGs and RQs, and our current understanding have been reviewed recently \citep[e.g.][]{Blandford2019,hardcastlereview20,Saikia22}. 
RLAGN with an overall projected linear size greater than 0.7 megaparsec (Mpc) are defined as giant radio sources (GRSs); those   associated with galaxies and quasars are referred to as giant radio galaxies (GRGs) and giant radio quasars (GRQs), respectively. They are quite rare compared with the RLAGN population, as shown in recent studies \citep[e.g.][]{PDLOTSS,Kuzmicz2018,sagan1,Delhaize21}. They are thought to grow in underdense environments and/or are supplied with continuous jet power for a long duration ($\sim$\,10$^{8}$~yr), thereby allowing them to scale larger distances. Efforts to constrain models explaining the giant nature of GRSs are still ongoing \citep{gk89,ravi96,Saripalli2005,hardcastle19,sagan1,sagan2,bruni20,KSJ2021,sagan3}. %More discussion on the above can be found in \citet{sagan1}, 
The nature of GRSs has been reviewed in \citet{Komberg09} and more recently in \citet{DSM22}.

The last few years have seen a resurgence in studies of GRSs owing to the new sensitive surveys at radio wavelengths, such as  the LOFAR Two Metre Sky Survey  \citep[LoTSS;][]{lotssshimwell}, which is largely complemented by optical spectroscopic surveys like the Sloan Digital Sky Survey (SDSS; \citealt{sdss00,sdssdr14}).

In the course of identifying GRSs and carrying out detailed multi-wavelength studies of GRSs under our project SAGAN\footnote{\url{https://sites.google.com/site/anantasakyatta/sagan}} \citep{D17,sagan1}, we have discovered a peculiar giant radio galaxy J223301+131502 (GRG-J2233+1315 for short) from the NRAO VLA SKY SURVEY (NVSS; \citealt{nvss}), as reported in \citet{D17}. 
In the current paper we present our results on GRG-J2233+1315 based on our optical and radio observations. The observational details and data analyses are described in Sect.\ \ref{sec:oba}.

GRG-J2233+1315 is hosted by the galaxy SDSSJ223301.30+131502.5, which exhibits a large diffuse stellar halo with apparent SDSS r-band magnitude (m$_{\rm r}$) of 15.21. It is classified as an S0 a-type galaxy \citep{HYPERLEDA} with an isophotal diameter of (logD$_{25}$) of 0.77$\pm$0.06. The stellar mass of the galaxy is $\sim$\,1.62\,$\times$\,10$^{11}$\,$\rm M_{\odot}$ \citep{Lin18}.

It  has also been identified as the brightest cluster galaxy (BCG) by a number of catalogues using different methods (e.g. red sequence and friend-of-friend algorithm):  the MaxBCG Catalog of 13,823 Galaxy Clusters from the Sloan Digital Sky Survey \citep{Koester07}, catalogued as MaxBCG338.25543+13.25070; the GMBCG Galaxy Cluster Catalog of 55,424 Rich Clusters from SDSS DR7 \citep{haokoester10}, catalogued as J338.25544+13.25070; and the WHL cluster catalogue \citep{whl12} of 132,684 galaxy clusters from SDSS-III data, catalogued as WHLJ223301.3+131503.
Although the methods used by \citet{Koester07} and \citet{haokoester10}  are similar to some extent, the method used by \citep{whl12} is different. In addition, the datasets used in the three above-mentioned  catalogues are three different data releases of SDSS photometric galaxies.
Hence, it is evident that GRG-J2233+1315 is residing in a dense cluster environment, which is in contrast to the suggestions that GRGs reside in  sparse environments \citep{mack98,ravi08,malarecki15}. In this paper we use the word dense to denote a higher-density environment than that of the general intergalactic medium or that associated with a field galaxy.

In the reporting paper of GRG-J2233+1315, \citet{D17} used the photometric redshift ($z$) of $\sim$\,0.093 from SDSS to estimate the total projected linear size of  GRG-J2233+1315 to be $\sim$\,1.71 Mpc. Based on the only available radio data of NVSS at 1.4~GHz, \citet{D17} observed a bright region, possibly part of the GRG on the western side of the host galaxy;   it was thought to be a `knot' in the jet that is unresolved.

The important issue is to examine the growth of such GRGs, which have managed to grow to megaparsec lengths despite being in a cluster environment. The combination of radio and X-ray data of such sources will enable us to study the intergalactic medium (IGM) as well as the intracluster medium (ICM). The interaction of the radio plasma with the hot ICM can lead to the formation of cavities, which show up as depressions in the X-ray surface brightness maps \citep{birzan04,McNamara05}. Research still continues on how these radio plasma features influence the ICM as they could be supplying a dynamically important amount of heat and/or magnetic field to the ICM \citep{Kronberg01,quilis01,brugenanture02,fabian02,heinz02,McNamara07}.
Hence, a GRG with 1.83 Mpc size, residing in a cluster environment with peculiar radio morphology consisting of bright unresolved features and diffuse lobes warrants a detailed study using multi-frequency radio data.

\section{Observations and data analyses} \label{sec:oba}
In this section we provide the details of the optical and radio observations and their corresponding data analyses for GRG-J2233+1315. The radio subsection is further divided into two sub-subsections (GMRT and LOFAR).

\subsection{Optical}\label{sec:wht}
An accurate redshift of the GRG is useful to estimate its properties, and hence efforts were made to update the SDSS photometric redshift of 0.093 (reported  in \citealt{D17}) with that of spectroscopic observations. The optical spectrum of the host galaxy (SDSS J223301.30+131502.5) was obtained using    long-slit spectroscopy at the 4.2m~William Herschel Telescope (WHT), located at La Palma, Canary Islands (Spain). The observations were carried out on 27 June~2017 under the programme ID  W17AN014, using the Intermediate dispersion Spectrograph and Imaging System (ISIS\footnote{\url{http://www.ing.iac.es/astronomy/instruments/isis/index.html}}) instrument with  3\arcmin ~slit length and    1.5\arcsec\ slit width. Further details of the observations and data reduction are presented in Section 2.2.1 of \citet{Saxena19}.

We   used the standard IAU conversion from air to vacuum wavelengths \citep{Vac2air-M91}.
The  centre, full width at half maximum (FWHM) and error on FWHM for emission lines of the object, and the lamp spectra were measured using the Gaussian fitting routines. The instrumental broadening of grism 7 is 6.78\AA. We   calculated the intrinsic FWHM using instrumental broadening. The optical spectrum obtained from WHT in two bands (red and blue) are combined and normalised using SDSS photometric flux. Briefly, we first generate a synthetic spectrum of the galaxy using SDSS photometric flux from the $ugriz$ bands. The synthetic spectrum is then fitted with a higher-order polynomial to get the continuum of the galaxy. This continuum is used to normalise the combined optical spectrum obtained from the WHT. We  calculated the redshift from the observed wavelengths, and the error in redshift was    calculated using corrected FWHM. 

\begin{figure*}
\centering
\includegraphics[scale=0.26]{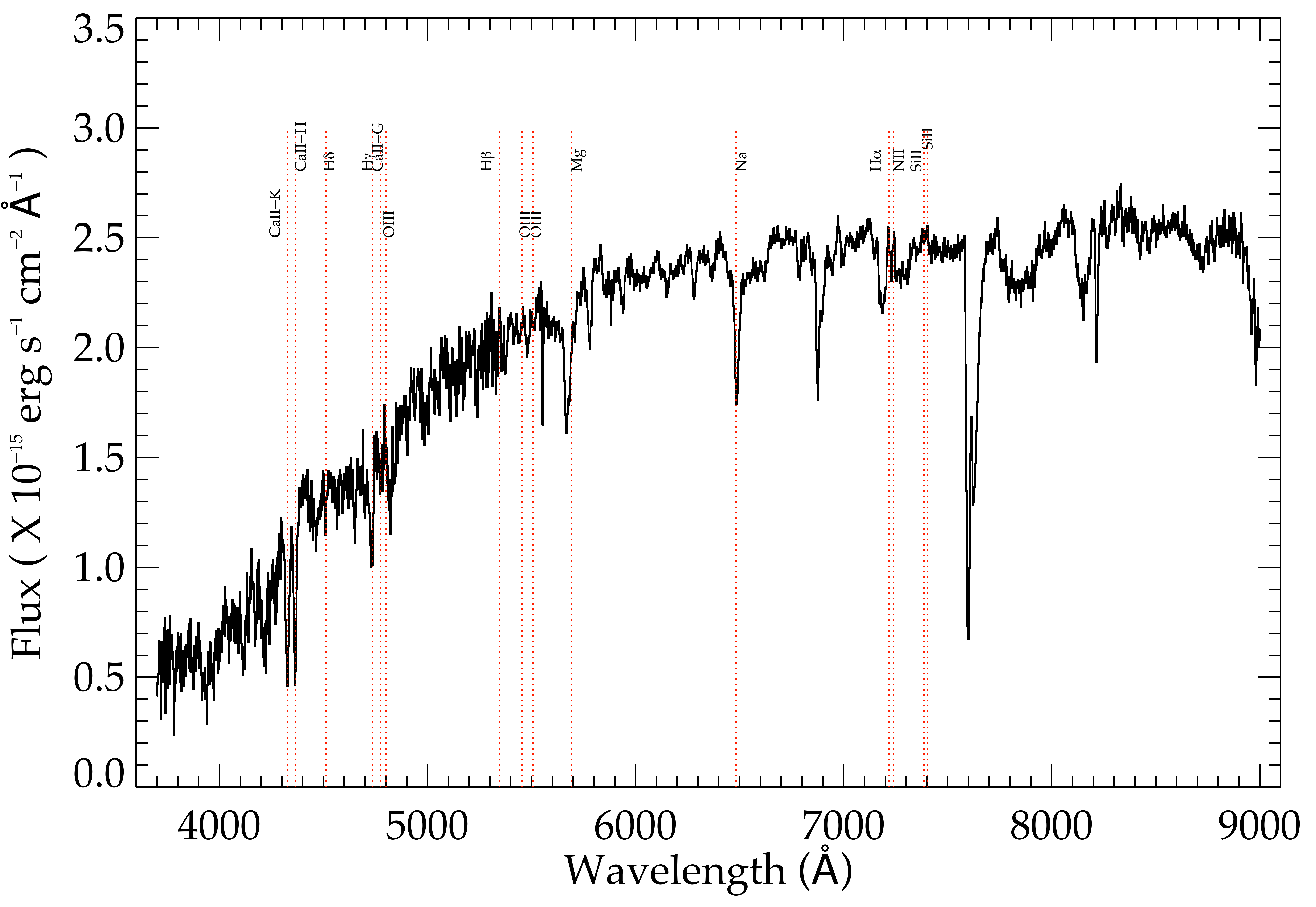}
\caption{Plot of   4.2m WHT ISIS combined blue and red arm  spectrum of   GRG-J2233+1315. The identified emission and absorption lines in the spectrum are marked with red dashed vertical lines. The redshift ($z$) of the galaxy is 0.09956 (see  Sect.\ \ref{sec:wht} for more details). The unmarked deep absorption lines are attributed to the atmosphere.}
\label{fig:whtspec}
\end{figure*}

The galaxy spectrum covering the wavelength range 3700--9000 \AA\ is shown in Fig.\ \ref{fig:whtspec}. 
The redshift ($z$) of the galaxy as identified from several prominent absorption and emission lines is 0.09956 $\pm$ 0.00383. More details are discussed in Sect.\ \ref{sec:opticalsed}.

\begin{table*}
\begin{small}

\renewcommand{\tabcolsep}{0.7mm}
\centering
\caption{Details of radio observations of GRG-J2233+1315. Observations at 1300, 612, and 323 MHz are with GMRT and at 144 MHz are with LOFAR. The primary calibrator refers to the flux density calibrator and the secondary is the phase calibrator. Observation date format is dd/mm/yyyy.}
\label{tab:obs}
\begin{tabular}{lcccccccccc}
\hline
Frequency & Project & Observation  & ObsID&Time on & Bandwidth & Channels & Integration & Correlations & Primary & Secondary \\
(MHz) & Code & Date & &Source (mins)  & MHz & & Time (secs) &  & Calibrator & Calibrator\\
(1) & (2) & (3) & (4) & (5) & (6) & (7) & (8) & (9) & (10)  & (11)\\
\hline
1300 & 31$\_$084 & 23/10/2016 &8940 &193 & 32 & 512 & 8 & RR $\&$ LL & 3C48 &2148+069\\
612 & 28$\_$020 & 17/05/2015 & 7785&225 & 32 & 512 & 8 & RR $\&$ LL & 3C48 & 2130+050\\
323 & 31$\_$084 & 03/11/2016 & 8962&349 & 32 & 512 & 8 & RR $\&$ LL & 3C48 &2212+018\\
144 & LC6$\_$024 & 05/09/2016 & 544937&480 & 49 & 231 & 1 & RR $\&$ LL & 3C196 & 3C196\\

\hline
\end{tabular}
\end{small}
\end{table*}

\subsection{Radio observations and data analyses} \label{sec:radioobs}
GRG-J2233+1315 was observed  at three frequencies with the Giant Metrewave Radio Telescope (GMRT) \citep{Swarup1991GMRT1,Swarup1991GMRT2} covering the frequency range from $\sim$\,300 to 1400 MHz (see Table\ \ref{tab:obs}).
We obtained radio observations at frequencies below 300 MHz with the LOFAR’s high band antennae (HBA) \citep{lofar}.
In the following we describe the observations and data analyses.

\subsubsection{GMRT observations and analyses} 
GRG-J2233+1315 was observed with GMRT at 323 MHz, 612 MHz and 1300 MHz during 2015-2016 (see Table\ \ref{tab:obs}) under the projects \texttt{28$\_{20}$} and \texttt{31$\_{84}$} (PI: Pratik Dabhade).
The raw visibility GMRT data of the three frequencies were analysed using the Source Peeling and Atmospheric Modelling (SPAM; \citealt{tgss_intema}) package. SPAM is based on the Astronomical Image Processing System (AIPS; \citealt{Greisen03}) and the  Python programming language. To access tasks in AIPS, a Python-based parseltongue interface \citep{Kettenis06} was used in SPAM. The analysis steps in SPAM start with the flux density and bandpass calibrations, which are derived from the respective calibrators after three cyclic iterations of flagging bad data including radio frequency interference (RFI) and calibrations based on models (e.g. \citealt{ScaifeHeald12}).
In addition, instrumental phase calibrations were also determined using the methods of \citet{Intema09}. For ionospheric calibration, direction-dependent gains were derived from the strong sources present in the field of view (FOV), which were then used to fit a time-variable phase screen over the whole array. This phase screen was used later during imaging to correct the entire FOV for effects caused by ionospheric phases.
The derived calibrations were transferred to the target data (GRG-J2233+1315). This was followed with imaging, which consisted of several cyclic iterations of self-calibration with phase only gain calibration initially using local sky model derived from surveys like NVSS and others. The iterations also involved flagging the remaining weaker RFI. Finally, wide-field facet-based imaging was carried out with the 
visibility weighting scheme of \texttt{ROBUST} -1 (AIPS) as it presents a finer point spread function. Three more iterations of self-calibrations were carried out with the addition of amplitude calibration in the final round. In each round, the process of peeling \citep{Noordam04} was performed (see \citealt{tgss_intema} for more details).

\subsubsection{LOFAR observations and analyses}
GRG-J2233+1315 was observed three times between September 2016 and March 2020, once in a dedicated eight-hour LOFAR observation (project LC6$\_$024, observation ID 544933, PI: Pratik Dabhade), and also as part of LoTSS where, if we account for the target being offset from the pointing centre, the effective integration time at the location of the target was  approximately eight hours  (project LT10$\_$010, co-observing project LC11$\_$009, and observation IDs 695799, 708070, 765089, 773699). These observations were all conducted in HBA Dual Inner Mode, and for this project we made use of 231 0.195\,MHz sub-bands between 120 and 169\,MHz. Two 10\,min calibrator observations were observed before and after each target observation. The basic details of these observations are summarised in  Table\ \ref{tab:obs}.

The data were processed using the PreFactor pipeline,\footnote{\url{https://github.com/lofar-astron/prefactor}} as described in \cite{vanWeeren2016}, \cite{Williams2016}, and \cite{deGasperin2019}, which corrects the data for direction-independent effects such as the bandpass, ionospheric Faraday rotation, clock offsets between different stations, and an offset between the  XX and YY phases. The data were then processed with the latest version of the LoTSS direction-dependent  calibration pipeline, the  DDF-pipeline,\footnote{\url{https://github.com/mhardcastle/ddf-pipeline}} which is summarised in Sect. 5.1 of \cite{lotssshimwell} and described in more detail in \citet{Tasse20}. This pipeline uses the calibration package kMS (\citealt{Tasse14} and \citealt{Smirnov15})  and imaging package DDFacet (\citealt{Tasse18}) to perform a direction-dependent self-calibration loop where calibration solutions are derived and gradually refined for 45 different directions spanning an 8$^{\circ} \times 8^{\circ}$ FOV and applied during the imaging. After each individual pointing is imaged we used the procedure described in \citet{vanWeeren20} to refine the solutions in the direction of our target and allow for each re-imaging. For each pointing this involved subtracting sources away from the target region before phase shifting all datasets to be centred on the target and then performing a direction-independent self calibration that makes use of WSClean \citep{Offringa14wsclean} and DPPP \citep{dppp}. 

\begin{table}[htbp]
\centering
\renewcommand{\arraystretch}{1.3} 

% \begin{minipage}{75mm}
\captionsetup{width=8.8cm}

\caption{Basic information about GRG. RA and Dec are right ascension and declination of the host galaxy (J2000 coordinates). $\rm S_{\upnu}$ is integrated flux density of the source and  $\rm P_{\upnu}$ is the radio power computed  at frequency $\upnu$. S$\rm _{1400 MHz}$ is measured from NVSS. $\alpha_{\rm int}$ is the integrated spectral index of the GRG from 144 MHz to 1400 MHz. Angular size refers to the largest angular size (LAS) of the GRG, as measured from the 3$\sigma$ contour of the LOFAR image at 144 MHz, which corresponds to 0.73 mJy beam$^{-1}$ . The flux density and corresponding power at 1300 MHz are from GMRT L-band observations. Here the $S_{\upnu}$ $\propto$ ${\upnu}^{- \alpha}$ convention is followed.}\label{tab:basic} 

\begin{tabular}{cc}
\hline
\hline
Properties & Values \\ 
\hline
RA & 22 33 01.30 \\ 
Dec & $+$13 15 02.52 \\ 
r$_{\rm band}$ (mag) & 15.04 \\
LAS ($\arcmin$) & 16.1 \\ 
Redshift ($z$) & 0.09956 $\pm$ 0.00383  \\ 
Projected Size (Mpc) & 1.83 \\ 
S$\rm _{1300 MHz}^{Core}$ (mJy) & 3.72 $\pm$ 0.84 \\ 
S$\rm _{1400 MHz}$ (mJy) & 231 $\pm$ 7.8 \\ 
S$\rm _{612 MHz}$ (mJy) & 479 $\pm$ 24.3 \\ 
S$\rm _{323 MHz}$ (mJy) & 835 $\pm$ 83.5 \\ 
S$\rm _{144 MHz}$ (mJy) & 1487 $\pm$ 297.4 \\ 
P$\rm _{1300 MHz}^{Core}$ ($\times$ 10$^{22}$ W Hz$^{-1}$) & 9.30 $\pm$ 2.20 \\ 
P$\rm _{1400 MHz}$ ($\times$ 10$^{25}$ W Hz$^{-1}$) & 0.60 $\pm$ 0.05 \\ 
P$\rm _{612 MHz}$ ($\times$ 10$^{25}$ W Hz$^{-1}$) & 1.3 $\pm$ 0.12 \\ 
P$\rm _{323 MHz}$ ($\times$ 10$^{25}$  W Hz$^{-1}$) & 2.2 $\pm$ 0.30 \\ 
P$\rm _{144 MHz}$ ($\times$ 10$^{25}$  W Hz$^{-1}$) & 3.9 $\pm$ 0.82 \\
$\rm \alpha_{144}^{1400}$  & 0.82 $\pm$0.18 \\
$\rm Q_{Jet}$ (erg s$^{-1}$) & 1.3 $\times$ 10$^{43}$  \\

\hline
\end{tabular}
% \end{minipage}
\end{table}

\begin{table*}
% \renewcommand{\tabcolsep}{1.0mm}
% %   \renewcommand{\tabrowsep}{1.7mm}
% \renewcommand{\arraystretch}{1.2} 
\setlength{\tabcolsep}{2pt}

 \centering
  \caption{Radio properties of GRG-J2233+1315, as measured from our observations from GMRT and LOFAR, along with archival NVSS data. WL and EL refer to the western and eastern lobe, respectively.}
  \label{tab:radioprop}
  \begin{tabular}{lcccccccccc}
    \hline
    Region & $\rm S_{1400 MHz}$ & $\rm P_{1400 MHz}$ & $\rm S_{612 MHz}$ & $\rm P_{612 MHz}$ & $\rm S_{323 MHz}$ & $\rm P_{323 MHz}$ & $\rm S_{144 MHz}$ & $\rm P_{144 MHz}$ & $\rm \alpha_{144}^{1400}$  \\
     & (mJy) & $\rm 10^{24}WHz^{-1}$ & (mJy) & $\rm 10^{25}WHz^{-1}$ & (mJy)  &  $\rm 10^{25}WHz^{-1}$ & (mJy)  & $\rm 10^{25}WHz^{-1}$ & \\
     (1) & (2) & (3) & (4) & (5) & (6) & (7) & (8) & (9) & (10) \\
    \hline \\

    WL & 30 $\pm$ 2.3 & 0.81 $\pm$ 0.08 & 90 $\pm$ 5 & 0.24 $\pm$ 0.02 & 177 $\pm$ 18 & 0.48 $\pm$ 0.06 & 344.3 $\pm$ 68.9& 0.93 $\pm$ 0.20 & 1.16$\pm$0.07  \\
    EL & 56.7 $\pm$ 3.1 & 1.5 $\pm$ 0.14 & 145.5 $\pm$ 7.8 & 0.39 $\pm$ 0.04 & 301.5 $\pm$ 30.3 &0.81 $\pm$ 0.10 & 559.6 $\pm$ 112.1 &1.50 $\pm$ 0.32 & 1.09$\pm$0.06  \\

    \hline
  \end{tabular}
 \end{table*}

\section{Results and discussion}
In this section we present the results obtained from the optical (Sect.\ \ref{sec:opticalsed}) and radio data analyses of  GRG-J2233+1315. We  present our discussion on the source in detail in the remaining subsections.

\subsection{Optical}\label{sec:opticalsed}
The optical spectrum of the host galaxy (Fig.\ \ref{fig:whtspec}) shows strong absorption lines from Ca H+K and Na, and lack of prominent emission lines, indicating that the galaxy is a passively evolving quiescent galaxy. 
Hence the galaxy appears to be a low-excitation GRG (LEGRG) having old stellar populations. 
\citet{Laing1994} classified high-excitation radio galaxies (HERGs) with the criterion of line flux ratios of [O{\sc iii}]
/ H$\boldsymbol{\alpha}>$ 0.2 and equivalent width (EW) of [O{\sc iii}] $> 3\AA$. \citet{Buttiglione2010} defined the excitation index (EI) comprising of line flux ratios of six lines ([O{\sc iii}], H$\boldsymbol{\beta}$, [N{\sc ii}], H$\boldsymbol{\alpha}$, [S{\sc ii}], and [O{\sc i}]) to classify HERGs and LERGs;  the sources with EI $>$ 0.95 were classified as HERGs. Both of the above-mentioned schemes have been widely used for identifying HERGs and LERGs \citep[e.g.][]{bh12rgs}.
As seen in Fig.\ \ref{fig:whtspec}, some of the lines (e.g. [O{\sc iii}] and H$\boldsymbol{\beta}$) are weak and/or not detected significantly, hence GRG-J2233+1315 can be classified as a LEGRG. As discussed in \citet{sagan1} and \citet{Simonte2022}, the LEGRGs are the dominant population in comparison with the  HEGRGs, analogously to the normal-sized radio sources \citep{bh12rgs,hardcastlenat}.

\subsection{Optical SED}
BAGPIPES \citep{Carnall18} is a Python-based package used to model the spectral energy distribution (SED) of the galaxy and estimate various physical properties such as stellar mass, dust attenuation, star formation rate (SFR), and  star formation history (SFH). To model the SED of GRG-J2233+1315, we  used SDSS photometric and WHT spectroscopic data in BAGPIPES. This   improves the precision of the flux calibration of the spectrum and the SED fitting. Since the redshift of the galaxy is well estimated from the optical spectrum, we adopted a Gaussian redshift prior with $\sigma=0.005$. We used the traditional approach for SED fitting with a simple exponentially declining SFH model. Exponentially declining SFHs are the most commonly applied SFH model. The model assumes that star formation jumps from zero to its maximum value at some time $T_0$, after which star formation declines exponentially on a timescale $\tau$,
\begin{equation}
   \rm SFR (t) \propto   exp(- \frac{t-T_0}{\tau})\, ;~ \rm{when}\, t>T_0,
\end{equation}
\begin{equation*}
 \rm   SFR (t) =   0 \, \rm{when}\, ;~ t<T_0.
\end{equation*}
We adopted the \citet{bruzal2003} library of single stellar populations to model the stellar emission of our galaxy with \citet{kroupa01} initial mass function (IMF).
We included the {\tt nebular} module to estimate the nebular emission from our galaxy.
The nebular emission model implemented in BAGPIPES is constructed following the methodology of \citet{Byler17}, using the latest
version of the Cloudy photoionization code \citep{Ferland17}.
The ionization parameter (U) determines the strength of the emission lines. For this model, we  kept U as a free parameter. The other default parameters were the solar abundances of \citet{Anders89},  ISM depletion factors, and the  helium and nitrogen scaling relations of \citet{Dopita2000}.
To account for dust attenuation caused to the stellar and nebular emission, we employed \citet{Calzetti00} {\tt dustatt\_modified\_starburst} module. We  used flat priors for all the input model parameters.

The stellar mass of the galaxy   obtained from our SED fitting model is $11.72^{+0.03}_{-0.02}$ $\times$ 10$^{10}$ M$_\odot$ (similar to that of \citealt{Lin18}). Moreover, the galaxy is likely to confine a moderate to high amount of dust inside given  the stellar dust attenuation, E(B-V)$_{\rm stellar}$ $\approx$ 0.23 mag. The age of the main stellar population (age$_{\rm main}$)  of the galaxy is $10.39^{+0.27}_{-0.36}$ Gyr and the SFR is in the range $10^{-3} - 10^{-3.5} \, \rm M_{\odot} yr^{-1}$ indicating an old quenched galaxy.

\subsection{Size, radio power, and jet kinetic power}
Using our radio observations with GMRT and LOFAR (see Sect.\ \ref{sec:radioobs}), we obtained deep and high-resolution images at 1300, 612, 323, and 144 MHz, as shown in Fig.\ \ref{fig:allims}. The deep GMRT 323 MHz and LOFAR 144 MHz images, which are more sensitive to diffuse large-scale structure, show the maximum extent of the source. The largest angular size (LAS) of 16.1\arcmin ~for emission above 3$\sigma$ was estimated from the LOFAR 144 MHz map, the lowest contour level at 3$\sigma$ being 0.73 mJy/beam. This yields a projected linear size of 1.83 Mpc. This was computed using the relation 
\begin{equation}
  \rm  D = \frac{\theta \times D_{c}}{(1+z)} \times \frac{\pi}{10800}
,\end{equation}
where $\theta$ is the LAS of the GRG on the sky in units of arcminutes (\arcmin), D$_{\rm c}$ (429.9 Mpc) is the comoving distance in Mpc, $z$ is the host galaxy's redshift, and D is the projected linear size in Mpc. In this paper the cosmological parameters from Planck are used (H$\rm _0$ = 67.8 km $\rm s^{-1}$ Mpc$^{-1}$, $\Omega\rm _m$ = 0.308, $\Omega\rm _{\Lambda}$ = 0.692; \citealt{2016A&A...594A..13P}; scale at $z$=0.09956 is 1.896 kpc/\arcsec). 

In Table\ \ref{tab:basic}, we  provide the integrated flux densities and their corresponding radio powers at the given frequencies. The CASA \texttt{imview} program was used to manually select the extended emission of the GRG in order to extract the flux densities. The flux densities of unrelated sources were  subtracted out. The core flux density was estimated using the GMRT 1300 MHz map using the CASA task \texttt{imfit}.

The radio power was computed using the formula $\rm P_{\nu}  = 4\pi D_L^{2}S_{\nu} (1+z)^{\alpha - 1}$, where $\rm D_{\rm L}$ is the luminosity distance (estimated using the redshift, $z$), $\rm S_ {\rm\nu}$ is the measured flux density at frequency $\rm \nu$,  $(1+z)^{\rm \alpha - 1}$ is the standard \textit{k}-correction term, and $\rm \alpha$ is the radio spectral index.
The total radio luminosity of the source at 144 MHz is $3.9 \times 10^{25}$ W Hz$^{-1}$ which is below the traditional lower limit for FRII sources, which is $\sim 10^{26}$ W Hz$^{-1}$ at 150 MHz \citep{mingo19}. However, recent LOFAR observations have demonstrated the existence of a significant population of FRII sources below this luminosity \citep{mingo19,Mingo22}.

The jet kinetic power ($\rm Q_{Jet}$) was estimated according to the scheme based on simulation-based analytical model given by \citet{Qjet_Hardcastle}, who has shown it to be a good estimator using low-frequency radio data. We estimate $\rm Q_{Jet}$ to be 1.3 $\times~10^{43}$~erg~s$^{-1}$, which is on the low side when compared to powerful FRII radio galaxies and quasars, which have $\rm Q_{Jet}$ usually above $\sim$\,10$^{45}$~erg~s$^{-1}$ \citep{Godfrey13}. As per Figure 3 ($\rm Q_{Jet}$ as a function of 150 MHz radio luminosity) of \citet{Godfrey13}, GRG-J2233+1315 lies very close to the division between FRI and FRII radio galaxies.

%%%%%%%%%%%%%%%%%%%%%%%%%%%
\begin{table}[htbp]
 \centering
\setlength{\tabcolsep}{2.7pt}

% \begin{minipage}{70mm}
\captionsetup{width=8.8cm}
\caption{Information of images shown in Fig.\ \ref{fig:allims}. The asterisk (*)  indicates a high-resolution image from LOFAR. }\label{tab:radioimtab}
\begin{tabular}{lccc}
\hline
Image & Frequency & RMS & Resolution \\ 
&(MHz) & (mJy~beam$^{-1}$) & (\arcsec $\times$ \arcsec, \deg) \\ 

\hline
a&1300 & 0.013 & 2.5\arcsec~$\times$~2.1\arcsec,~62.3$^{\circ}$\\
b&612 & 0.039 & 7.1\arcsec~$\times$~6.8\arcsec,~22.3$^{\circ}$\\
Kink&612 & 0.031 & 5.1\arcsec~$\times$~4.5\arcsec,~31.5$^{\circ}$\\
c&323 & 0.081 & 12.1\arcsec~$\times$~8.8\arcsec,~51.9$^{\circ}$\\
d&144 & 0.242 & 20.0\arcsec~$\times$~20.0\arcsec,~90.0$^{\circ}$\\
d$^{*}$&144 &0.145 & 9.9\arcsec~$\times$~5.5\arcsec,~80.9$^{\circ}$ \\

\hline
\end{tabular}
% \end{minipage}
\end{table}

%%%%%%%%%%%%%%%%%%%%%%%%%%%

\subsection{Radio morphology}\label{sec:rmorph}
\begin{figure*}
\centering
\includegraphics[scale=0.21]{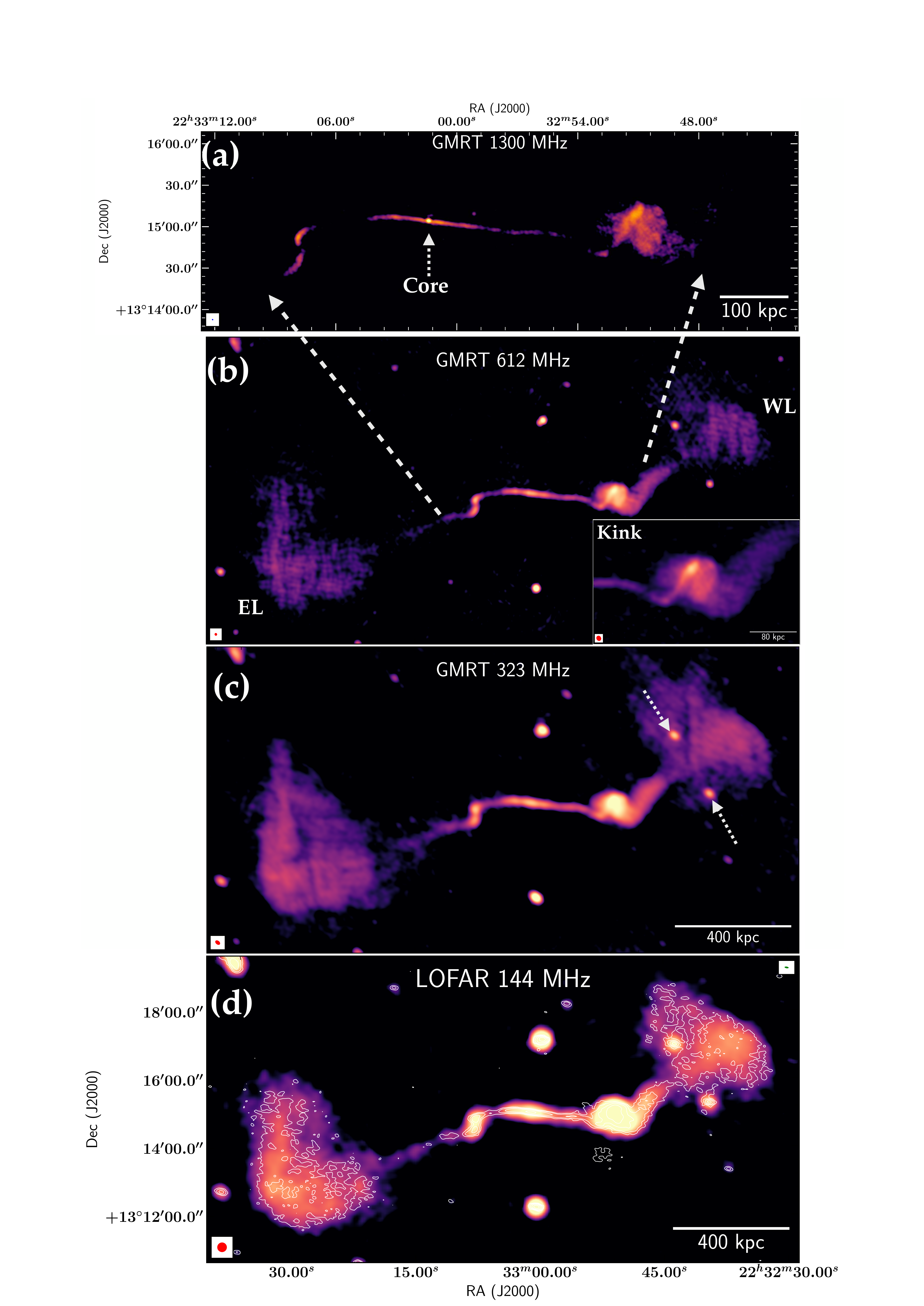}
\caption{{Radio images of GRG-J2233+1315 at 1300 (a), 612 (b), 323 (c), and 144 MHz (d). The RMS and resolution information of the images are provided in Table\ \ref{tab:radioimtab}.
The white box at the bottom left represents the beam. In  panel (b) an additional zoomed-in image of the kink feature of the GRG is shown (see inset at bottom right).
The two arrows in panel (c) show two unrelated sources. In panel (d) the  LOFAR 144 MHz image is shown with two resolutions. The higher resolution  is represented by contours, whose beam is shown in the top right corner of the panel. The radio contours in panel (d) are drawn with six levels, which are chosen by equally (in log scale) dividing the data value range above 3$\sigma$ (local rms of the map).}}
\label{fig:allims}
\end{figure*}

Using the NVSS radio map of GRG-J2233+1315, \citet{D17} reported a bright unresolved feature on the western side of the host galaxy, which was thought to be a knot in the jet. Using our deep and high-resolution observations with GMRT and LOFAR, we  resolved the overall structure of  GRG-J2233+1315 in detail. Our multi-frequency radio maps from GMRT and LOFAR  as seen in Fig.\ \ref{fig:allims} (image details in Table\ \ref{tab:radioimtab}), reveal a thin collimated jet emanating from the radio core\footnote{The radio core is detected only in the GMRT 1300 MHz map.} transforming into a $\sim$\,100\,kpc structure shaped like an upper case omega ($\Omega$), which we refer to as the kink on the western side. It is most likely that the kink feature is also present on the eastern side in some form of symmetry. While the
kink on the western side of the jet is prominently seen, the feature or the possible kink (or `twist')  on the eastern side of the jet does not show the  same morphology. It appears that  these two features are in different planes, and that  only one of them is clearly
observed due to projection effects.
The radio images of all frequencies in our study show this highly complex and rare feature in the structure of the GRG.
Beyond the omega-shaped ($\Omega$) kink feature, the jet appears to re-collimate again having a greater width than before and connecting to the inner edge of the mushroom-shaped lobe.
The two-point sources embedded in the western lobe (WL) (Fig.\ \ref{fig:allims} c) are unrelated sources and their flux densities were subtracted from the total flux density of the lobe.

Our highest resolution (2.5\arcsec~$\times$~2.1\arcsec) radio map at 1300 MHz from GMRT reveals the radio core, which coincides with the galaxy SDSS\,J223301.30+131502.5 and confirms its host nature.
On the eastern side  we observe a twisted jet feature that leads towards the lobe. The jet terminating in the eastern lobe (EL) is fainter, possibly due to Doppler dimming, compared to the western side. The full jet connection to the lobe can be seen only in the GMRT 323 MHz  and LOFAR 144 MHz maps.

Overall, the radio morphology of GRG-J2233+1315 (hearafter    Barbell GRG) resembles a barbell, the weightlifting equipment. Even though the Barbell GRG shows two lobes, as do  typical FRII RGs, it lacks the presence of   hotspots, and hence it is difficult to categorise the source strictly in the classical definition of FRI or FRII morphology. However, the presence of collimated jets connecting to the lobes is the same as seen in typical RGs or GRGs of FRII type. The study of \citet{mingo19} using LoTSS has clearly shown populations of FRI and FRII radio sources not obeying the canonical FRI--FRII radio power or luminosity break of 10$^{26}$ W~Hz$^{-1}$ at 150 MHz.

The Barbell GRG has   a conspicuous resemblance to another cluster centre source, 4C35.06 \citep{Riley75,Biju14}. As shown clearly in the particular study by \citet{bijua407}, the radio source 4C35.06
resides at the centre of the Abell 407 cluster with very complex radio morphology and perplexing scenario of nine merging galaxies. Hence the source was named `Zwicky's Nonet' by \citet{bijua407}, where one of the nine merging galaxies is the actual host of the radio galaxy. As seen in Figure 5 of \citet{bijua407}, features labelled   `A1' and `D2' are very similar to the WL and EL of the Barbell GRG along with the narrow re-collimated jet leading to the steep spectrum diffuse lobes.

\begin{figure}
\centering
\includegraphics[scale=0.31]{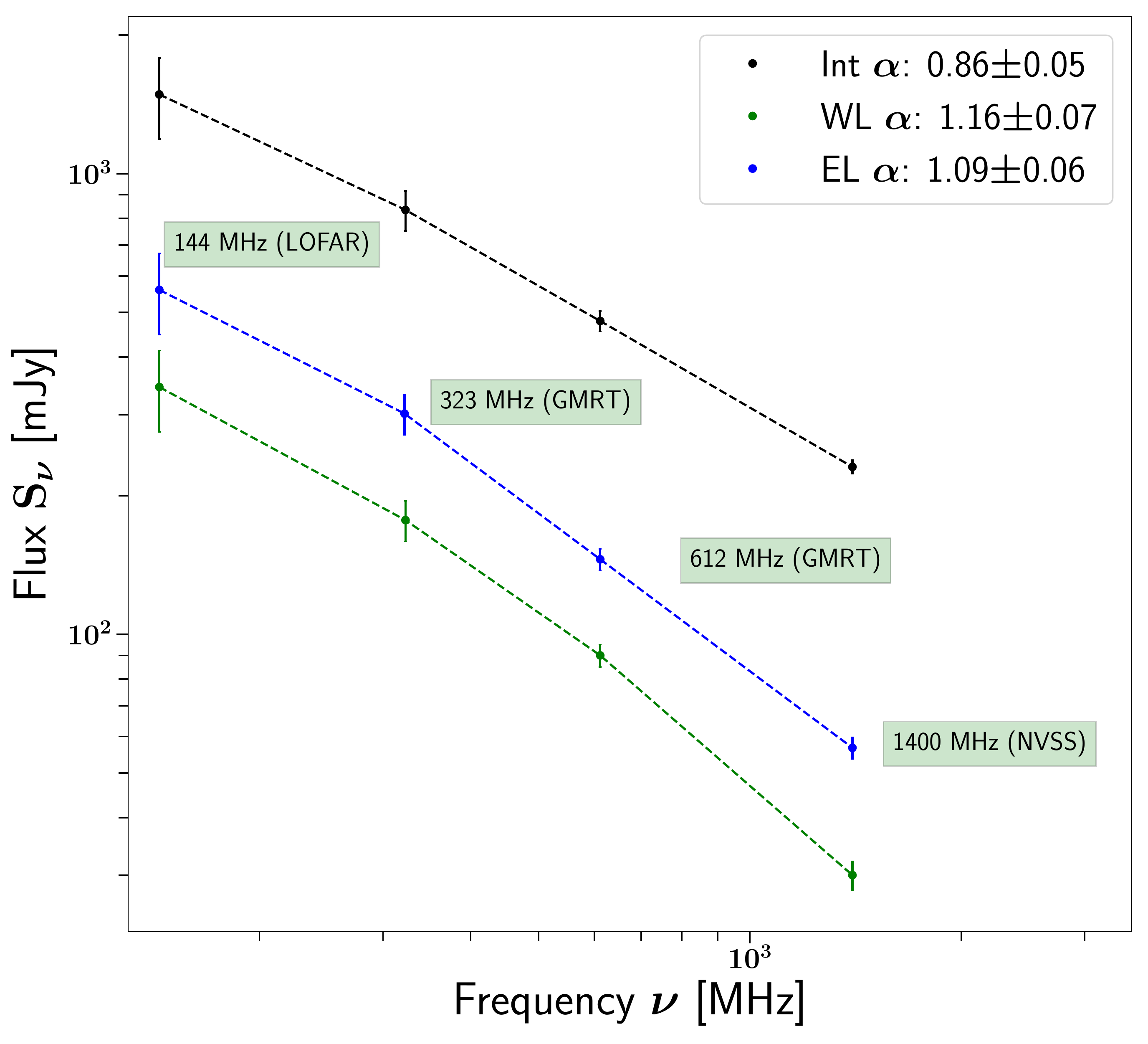}
\caption{Spectral index plot of western  and eastern lobes (WL and EL)  along with the full source (Int), where the 1400 MHz flux density was obtained from archival NVSS data. A possible break frequency is seen around $\sim$\,323~MHz.}
% Here, the $S_{\nu}$ $\propto$ ${\nu}^{ \alpha}$ is used. The rest of the paper follows, $S_{\nu}$ $\propto$ ${\nu}^{- \alpha}$ convention.}
\label{fig:SIPLOT}
\end{figure}

\begin{figure*}
\centering
\includegraphics[scale=0.44]{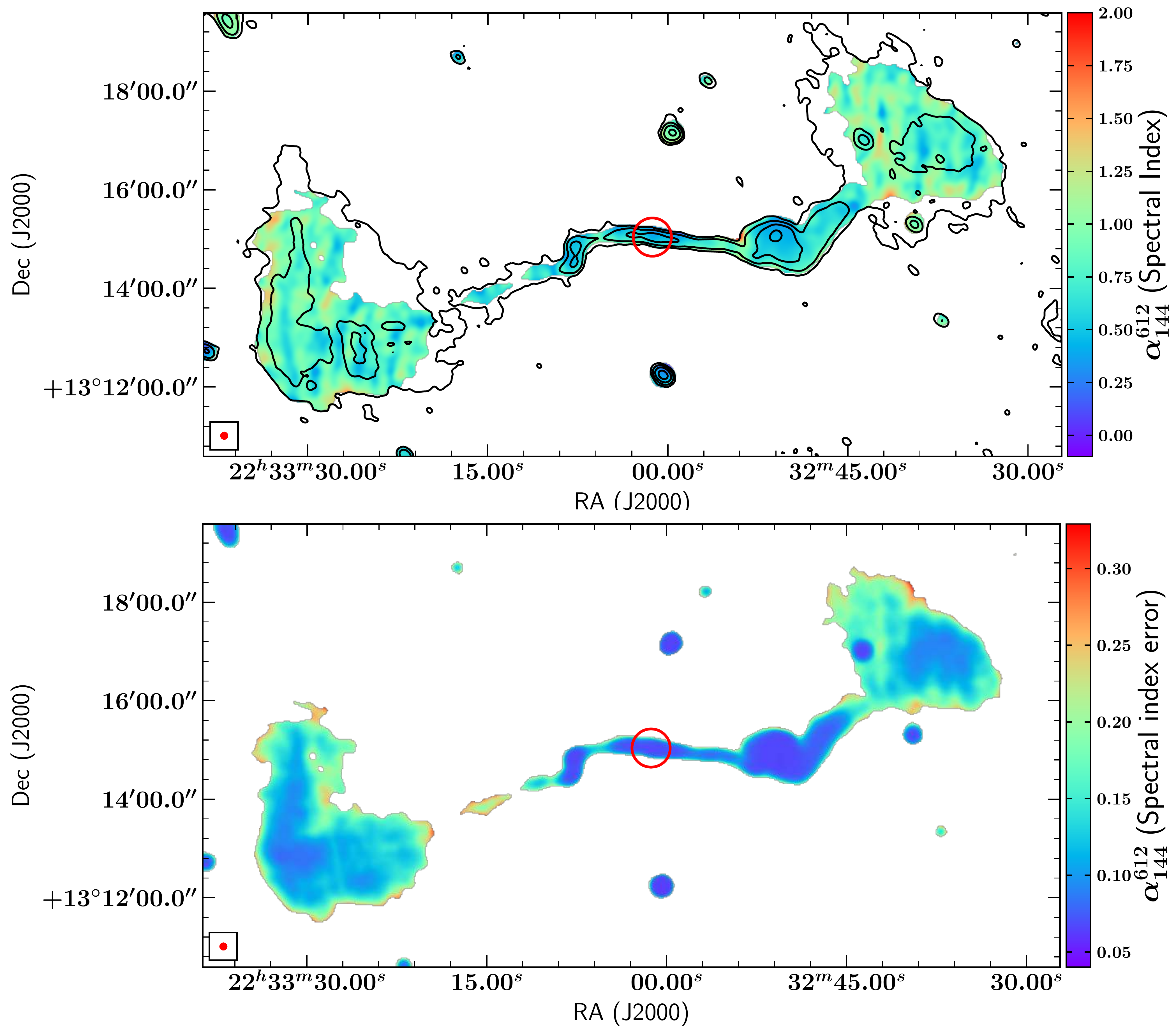}
\caption{Figure shows three frequency spectral index map and its uncertainties.
Upper panel: Spectral index map made using LOFAR 144 MHz, GMRT 323 MHz, and GMRT 612 MHz maps (described in Sect.\ \ref{sec:spinmaking}). The GMRT 323 MHz contours are overplotted in black   with six levels spaced in log scale with the lowest level of 3$\sigma$ (Table\ \ref{tab:radioimtab}).
The red   circle indicates the location of the radio core or host galaxy. In the  bottom left corner, the red dot   represents the beam of 13\arcsec $\times$ 13\arcsec. Lower panel:  Spectral index error map.}
\label{fig:simap}
\end{figure*}

\begin{figure*}
\centering
\includegraphics[scale=0.4]{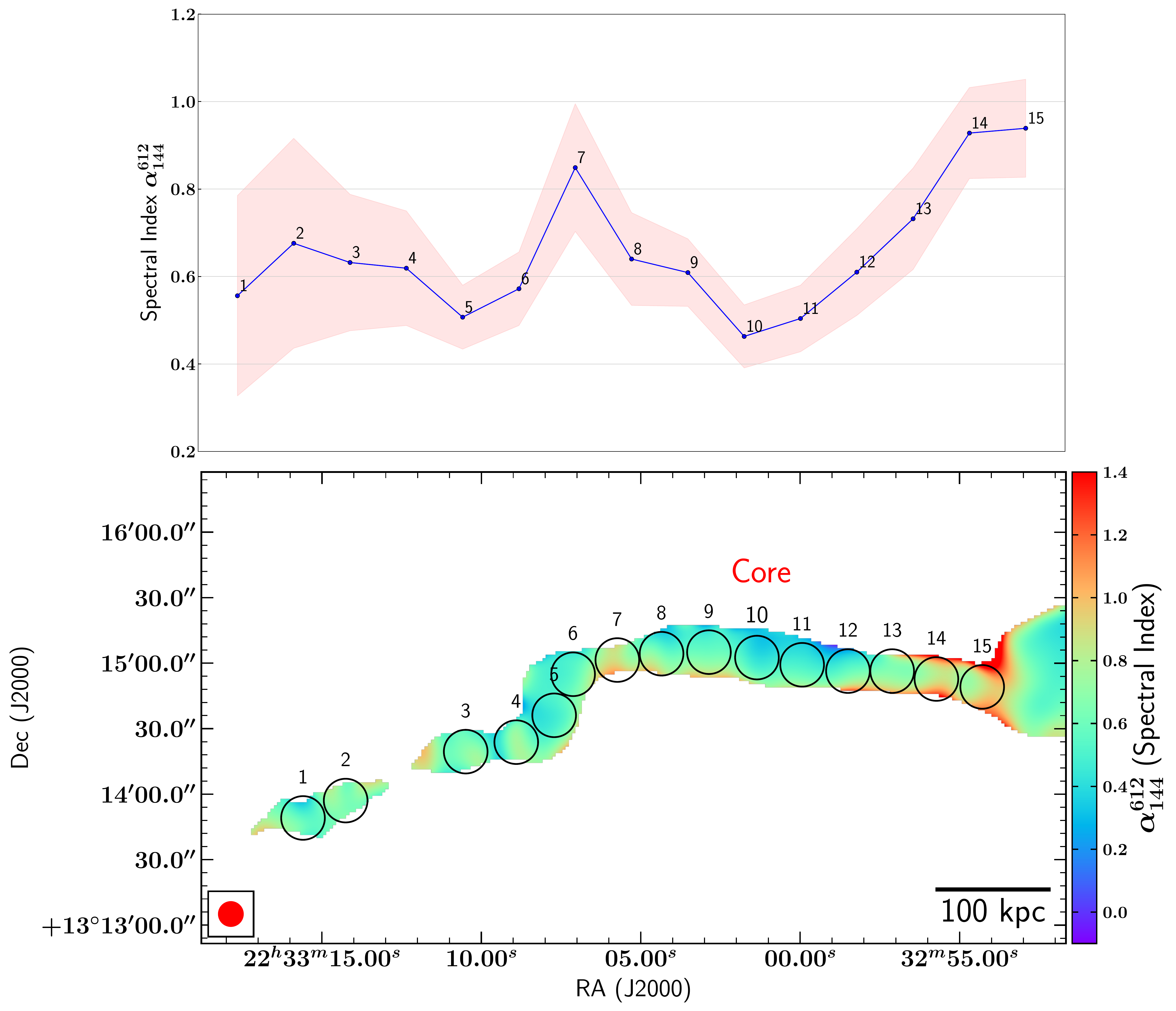}
\caption{The figure shows spectral index variation in the jets of the source.
Upper panel: Jet spectral index profile with respect to the marked regions seen in the lower plot. The shaded region indicates the uncertainties on the measurements.
Lower panel:  Zoomed-in portion of Fig.\ \ref{fig:simap} highlighting the core and jet.  The numbered regions are labelled.}
\label{fig:jetspec}
\end{figure*}

\subsection{Spectral index ($\alpha$)} \label{sec:spinmaking}
The integrated spectral index ($\rm S_{\upnu}$ $\propto$ ${\upnu}^{- \alpha}$) of the whole Barbell GRG and both lobes from 144 to 1400 MHz can be seen in Fig.\ \ref{fig:SIPLOT}, where the lobes are steeper ($\alpha$~$\sim$\,1). 

In Fig.\ \ref{fig:simap} we see the three-point spectral index map of the Barbell GRG made using LOFAR 144\,MHz, GMRT 323\,MHz, and GMRT 612\,MHz data at a resolution of 13\arcsec.
Here, the frequency coverage on the higher side was restricted to 612 MHz due to lack of similar resolution maps at $\sim$\,1400 MHz which recover the entire source. The NVSS map with its coarser resolution of 45$\arcsec$ does not resolve the jets and kink feature well. The GMRT 1300 MHz map provides the highest resolution of $\sim$\,2\arcsec, but although we tried imaging the source by using the short spacings and tapering the data, it did not recover the whole source. Therefore, in order to obtain spectral index information of the whole source with an angular resolution of 13\arcsec, the NVSS and GMRT 1300 MHz data were not used.

The steps followed for making the three-frequency spectral index maps are based on the procedure given in \citet{duy17}. Firstly, the individual datasets (maps presented in Fig.\ \ref{fig:allims}) were further re-imaged with a similar UV range (0.2$k\lambda$ to 50$k\lambda$) and weighting schemes with UV tapering to obtain maps at a common resolution of 13\arcsec. From each map, only those pixels with a brightness of greater than three times  the rms noise were used for the spectral index calculations and images. The GMRT images were aligned with respect to the LOFAR image along with regridding all images to a common pixelisation. For the process of image aligning, we fitted the compact point sources with 2D Gaussian function to find their accurate locations, and then the average displacement was determined between the three images.
Likewise, the GMRT images were appropriately shifted along the RA and Dec axes.

The spectral index map   in Fig.\ \ref{fig:simap} clearly shows the flatness ($\sim$\,0.5) of the radio core and the kink. Both the lobes show steep ($\sim$\,1) spectral index indicating older plasma (Table\ \ref{tab:radioprop}).

\begin{figure*}
\centering
\includegraphics[scale=0.22]{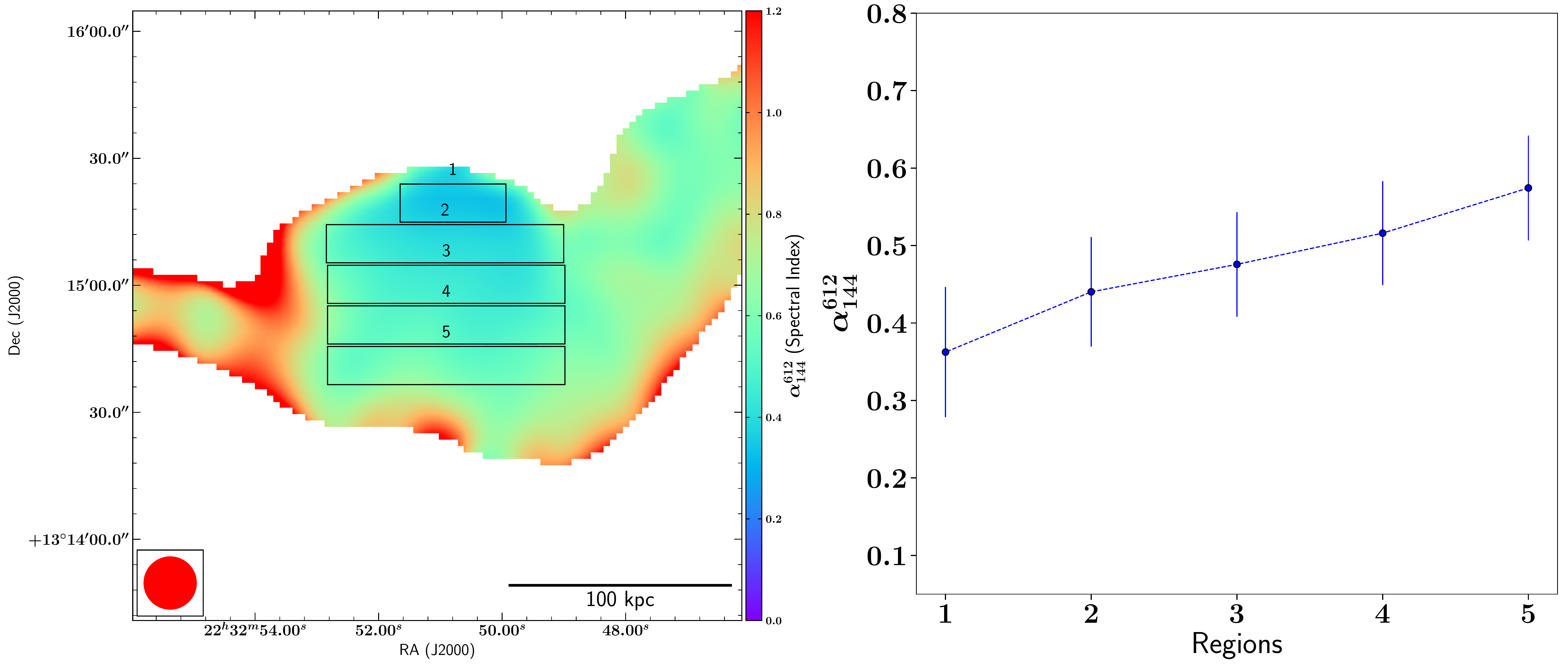}
\caption{ Spectral index variation in the kink.
Left: Kink region of the Barbell GRG from Fig.\ \ref{fig:simap}, with five box regions, where the first box is the top-most part of the kink. Right: Spectral index gradient of the kink  from regions 1 to 5. The top box, or region 1, has the flattest spectral index.}
\label{fig:kinkprspec}
\end{figure*}

\subsection{Spectral ageing and magnetic field} \label{ageingmag}
One of the ways to estimate the life span of radio galaxies is through interpreting the radio spectrum ($S_{\upnu}$ $\propto$ ${\upnu}^{- \alpha}$) of their electron population, often known as the spectral age of the source. 

The shape of the spectra is influenced primarily by two factors, the injection of energetic electrons and the electron population undergoing synchrotron and inverse-Compton losses. Energy losses for high-energy electrons are indicated by a steepening of the spectrum beyond the break frequency ($\rm \upnu_{b}$). It becomes steeper when the electrons are not rejuvenated or re-accelerated after the pause or end of AGN activity. The low-frequency (MHz) radio observations play an important role in determining the break in the power-law that occurs due to radiative losses. As more time elapses, the $\rm \upnu_{b}$ moves to lower radio frequencies \citep{Kardashev62,Pacholczyk,JP73}.
With the knowledge of the break frequency and the magnetic field of the source, the spectral age ($\rm \tau_{sp}$ in megayears,  Myr) can be calculated using the   formula \citep{vanderLaan1969,Leahy1991book} :

\begin{equation}
\rm \tau_{sp}=1590\frac{B^{\rm 0.5}}{(B^2+B_{\rm IC}^2)\sqrt{\upnu_{\rm b}(1+z)}} \\ ,
\label{eqtime}
\end{equation}where $\rm B_{IC}$ = 3.18$(1+z)^2$ is the magnetic field strength equivalent to the cosmic microwave background radiation and B is the magnetic field strength in various regions of the source,  both   in units of  $\muup$G. Here $z$ is the redshift of the source and $\rm \upnu_{b}$ is the spectral break frequency in GHz above which the radio spectrum steepens from the initial power-law spectrum.  In the simple spectral ageing model, the magnetic field is assumed to be constant in a given region throughout the energy loss process. It is also assumed that there is no energy loss due to the expansion process and the particles have a constant power-law energy spectrum. To determine the spectral ages of  the WL and EL of the Barbell GRG, a cylindrical geometry is assumed, and the sizes ($d$) of the specified regions are computed from the lowest frequency map of LOFAR 144 MHz. 
% For all measurements, filling factor of unity is considered.
The magnetic fields in the corresponding regions are calculated using the classical formalism \citep{Miley} and the revised formalism \citep{Brunetti1997,Govoni04,Beck_Krause}  by assuming minimum energy conditions. The following formulae have been used for calculations :
\begin{equation}\label{eq:umin}
\begin{split}
u_{\rm min}\Big[{\rm \frac{erg}{cm^3}}\Big] = \xi(\alpha, \nu_1, \nu_2)(1 + k)^{4/7}(\nu_{0}\,[{\rm MHz}])^{-4\alpha/7} \\
          \times~(1 + z)^{(12-4\alpha)/7}\Big(I_{0}\,[{\rm \frac{mJy}{arcsec^2}}]\Big)^{4/7}(d\,{[{\rm kpc}]})^{-4/7}
\end{split},
\end{equation}

\begin{equation}\label{eq:beq}
\rm B_{eq} [G] = \Big(\frac{24\pi}{7}\,u_{\rm min}\Big)^{1/2},
\end{equation}

\begin{equation}
 \rm B'_{eq} [G] = 1.1 \gamma^{\frac{1-2\alpha}{3+\alpha}}_{\mathrm{min}} \times B_{eq}^{\frac{7}{2(3+\alpha)}}.
\end{equation}

Assuming a uniform magnetic field and an isotropic particle distribution, the minimum energy density ($u_{\rm min}$) is computed, where the parameter $\xi$  is a function of spectral index $\alpha$. Different values of $\xi$ are given in Table 1 in \citet{Govoni04}, and an appropriate value is chosen based on our spectral index estimates. In the classical formalism the spectrum is integrated from $\upnu_{1}$ to $\upnu_{2}$, usually from 10 MHz to 100 GHz, where the corresponding minimum Lorentz factor ($\gamma_{min}$) and maximum Lorentz factor ($\gamma_{max}$) depend on the magnetic field strength. We assume a $\gamma_{min}$ of 100 to estimate $\rm B'_{eq}$ \citep[e.g.][]{Isobe2015,Giacintucci2021}. The quantity $k$ represents the ratio of energies in relativistic protons to electrons (1 or 100), I$\rm _{0}$ is the surface brightness at the measuring frequency, and $d$ is the mean of the major and minor axes (source depth). In addition, we assume the filling factor to be 1.
The terms $\rm B_{\rm eq}$ and $\rm B'_{\rm eq}$ respectively represent the classical and revised equipartition magnetic fields, which are related to $u_{\rm min}$, as seen in equation~\ref{eq:beq}.

Based on our analysis, the equivalent magnetic field of the lobes is around 5~$\muup$G (Table\ \ref{Tab:Specage}) which is similar to that of other GRGs \citep{Ishwara1999,Konar2004,kronberg04,Harwood16}. In addition, when compared with RGs of similar radio power and located in galaxy cluster centres from the  \citet{birzan08}  sample, the magnetic field strengths are comparable.
It has been suggested that powerful radio galaxies, especially GRGs covering megaparsec-scale volumes are strong candidates for magnetising the intergalactic medium
\citep{Rees1987,Ruzmaikin1989,Jafelice1992,kronberg94,Kronberg01}. This largely depends on how powerful the RG or GRG is along with its environment and hence, its direct effect may vary with the source. The study by \citet{Jafelice1992}, which  models powerful radio galaxies (e.g. M84, 3C~465, and NGC~6251) based on observations,   estimates the intergalactic medium magnetic fields to be in the range  10$^{-6} - 10^{-8}$ G.

Considering the synchrotron radiation mechanism in which the electrons are losing energy, the spectral ages are defined as   the time progressed since the last acceleration of the particles. In addition to the above-mentioned radiative cooling losses, the depletion in particle and field energy in the radio lobe can also be caused by adiabatic expansion \citep{scheuer68}. This can lead to a reduction in the magnetic field and break frequency, causing the measured spectral age to surpass the actual age of the radio source.

As we have flux density measurements at only four frequencies between 144 and 1400 MHz, the spectral age estimates given here require confirmation from additional observations at both lower and higher frequencies. Lower-frequency measurements   give better constraints on the injection spectral indices.  The two-point spectral indices between 144 and 323 MHz for the EL and WL  are 0.78$\pm$0.28 and 0.82$\pm$0.28, respectively. The corresponding values between 323 and 1400 MHz are 1.14$\pm$0.07 and 1.24$\pm$0.08, respectively. This suggests that there may be a spectral break around 300 MHz, but  more data is required to confirm it.

Our results for the spectral ages of the Barbell GRG are presented in Table\ \ref{Tab:Specage}, where adopting the commonly used k-value of 1, average spectral ages of $\sim$\,186 Myr (classical) and $\sim$\,124 Myr (revised) were   obtained (assuming $\rm \upnu_b$ of 323 MHz) for the radio lobes. To match the spectral age with the dynamical age, the jet head speed would need to be $\sim$\,0.03c (considering the distance between the radio core and the WL and constant jet speed). On the other hand, the lobes appear to be relics with no hotspots. In this case, if the jets are from a more recent cycle of activity, the velocity of the jets could be significantly higher, as seen in detailed models of jets \citep[see review by][]{Saikia22}. 

The spectral age estimation suffers from some uncertainties, such as  inhomogeneity in the magnetic field and lack of accurate injection index values. It is often found that the spectral age differs from the dynamical age of the source, which mainly depends on the geometry and brightness of the radio source (e.g. \citealt{machalski09}). 

The spectral ages of both lobes for the Barbell GRG are  on the high side compared to RGs; however, past studies of GRGs on spectral ageing have yielded a wide range of values between $\sim$\,10\,Myr and 250 Myr \citep{mack98,Schoenmakers_spec,Konar2004,Konar2008,Jamrozy08,Shulevski19,cantwell20}. It is worth noting that the method for estimating spectral ages was not uniform across all the above-mentioned spectral ageing work on GRGs.

A more detailed spectral ageing analysis is possible with additional observations at higher frequencies which will be presented in future work.

%%%%%%%%%%%%%%%%%%%%%%%%%%%%%%%%%%%%%%%%%%%%%%%%%%%%%%%%%%%%%%%%%%%%%%%%%%
\begin{table*}
\begin{center}
\setlength{\tabcolsep}{5pt}

% \begin{minipage}{130mm}
\caption{Values of equipartition magnetic field and spectral age estimated for the entire source (see Sect.\ \ref{ageingmag}). Here, I$_{0}$ is the surface brightness at frequency $\nu_{0}$, which is taken as 144 MHz. The assumed break frequency ($\rm \upnu_{b}$). The terms with (cl) use the classical formalism, and those with (rev)   use the revised formalism.}
\label{sa}
\begin{tabular}{lcccccccccc}
%\toprule[1.5pt]
\hline
Region & k-value  & I$_{0}$& $\alpha$ & $d$ & \textit{u}$_{\rm min}$ &$\rm B_{eq}(cl)$   & $\rm B'_{eq}(rev)$ &($\rm \upnu_b$)& $\rm \tau_{sp}(cl)$ & $\rm \tau_{sp}(rev)$ \\  
  & & (mJy~arcsec$^{-2}$)  &  & (kpc)& $\times 10^{-12}$ (erg cm$^{-3}$) &$(\rm \muup G $) &( $\rm \muup G) $ & (MHz)  &(Myr) &  (Myr)\\\hline \hline

WL & 1 & 3.68 & 1.16 & 530 & 1.24 & 3.7  & 6.8  & 323 & 181 & 114 \\
EL & 1  &3.48 & 1.09 & 573 & 0.94 & 3.2  & 5.7  & 323 & 191 & 134 \\

 \hline

\end{tabular}
\label{Tab:Specage}
% \end{minipage}
\end{center}
\end{table*}
%%%%%%%%%%%%%%%%%%%%%%%%%%%%%%%%%%%%%%%%%%%%%%%%%%%%%%%%%%%%%%%%%%%%%%%%%%

\subsection{Jet and the kink}
The Barbell GRG shows a remarkably long collimated jet, where the measured projected length of the jet from core to the base of the prominent kink is $\sim$\,125\arcsec, which  corresponds to $\sim$\,237\,kpc. Radio galaxies exhibiting long collimated jets extending to hundreds of kiloparsec ($>$\,200\,kpc)  are very rare (e.g. NGC~315 and  HB13, \citealt{Jaegers87}; CGCG 049-033, \citealt{bagchi07}; and 4C~34.47, \citealt{hocuk10}).
However, unlike these sources, the Barbell GRG exhibits a counter-jet on the eastern side, which is rare in GRGs with FRII morphology (e.g. NGC~6251; discussed later in this section). In  panels (c) and (d) of Fig.\ \ref{fig:allims} we clearly see the counter-jet leading to the EL. As noted earlier, the jet has a prominent kink at a distance of 127.5\arcsec (241 kpc) from the core,  while the counter-jet has a weaker kink, separated from the core by 79.6\arcsec (151 kpc).

The transverse widths of the radio emission beyond the kink and twist suggest that they are more likely to be collimated structures extending the jets to the inner edges of the lobes. 
The well-collimated jets (as seen in Fig.\ \ref{fig:allims} and Fig.\ \ref{fig:simap}) are similar to those seen in FRII radio sources, although the outer lobes have no prominent hotspots. Therefore, it is quite possible that the jets belong to a new episode of jet activity,
while the outer mushroom-shaped lobes are relics of a previous 
episode. In addition, the kink and twist, which are not quite symmetric about the core, could be the  transition points where the
inner jets meet the inner dense core  edge of
the hot X-ray gas that usually surrounds the BCGs and can possibly lead to
plasma instabilities causing the observed peculiar features. 

There are many similarities between the Barbell GRG and  NGC~6251, which is also a $\sim$\,2\,Mpc GRG residing in a   denser environment with a $\sim$\,200\,kpc one-sided (north-western) jet \citep{Waggett77}. This jet is known as the blowtorch jet, and has been the subject of several studies  \citep[e.g.][]{Perley84,mack97,Evans05} ranging from radio to X-ray wavelengths. A counter-jet of $\sim$\,50\,kpc was detected by \citet{Perley84} using VLA, and recently better maps have been presented showing extended jet emission by \citet{cantwell20} using LOFAR and VLA.
The spectral ages ($\sim$\,200\,Myr)  are  similar for   NGC~6251 \citep{cantwell20} and Barbell GRG, which are     older   compared to other GRGs from the literature with measurements. 

The spectral index map made using LOFAR 144 MHz, GMRT 323 MHz, and GMRT 612 MHz maps is shown in Fig.\ \ref{fig:simap}, while the spectral index ($\alpha$) along the jet and the counter-jet is shown in Fig.\ \ref{fig:jetspec}. 
Although the jet on the eastern side exhibits larger errors than on the western side and variations along its length, the value of spectral index is consistent with a value of $\sim$\,0.6 along most of the jet. However, on the western side the spectral index appears to systematically increase from $\sim$\,0.5 to 0.9. While more sensitive multi-frequency observations would be helpful to further examine these trends, a nearly constant spectral index or a flattening of the spectral index as the jet travels outwards suggests the reacceleration of particles.

The radio jets produced near the central supermassive black holes are collimated by the magnetic field, and their collimation requires external confinement \citep{Tomimatsu1994,Beskin1998}. For the external confinement the gas pressure profile plays a vital role, and hence influences the jet acceleration and collimation along with the magnetic field configuration. For the jet in GRG NGC~6251, \citet{Perley84}   discussed a model for jet confinement at large scales due to thermal pressure of the environment and how its non-uniformity can affect the overall source morphology.
Hence, in the  Barbell GRG it is interesting to see a staggering $\sim$\,237\,kpc collimated jet propagating in a low-mass galaxy cluster environment with weak X-ray emission, although there was no detection in the ROentgen SATellite (ROSAT) sky survey. The absence of the gas pressure profile of this dense environment means that it is not possible to model the parameters of the jet.

We first consider the possible explanations for the kink and twist in the jet, and then briefly comment on the overall structure of the source.
The magnetohydrodynamic (MHD) processes govern the launch, acceleration, and collimation of astrophysical jets. Several works over the past three decades \citep{Hummel92,baum97,Feretti99,Lobanov01,Marshall01,Mertens16}  have shown the presence of peculiar features in the jets on subparsec to kiloparsec scales, which appear to be `wiggles' or kinks. However, almost all of them are on smaller scales ($\sim$\,10\,kpc), and not scales of $\sim$\,100\,kpc as in the case of the kink in the Barbell GRG.
The $\sim$\,100\,kpc omega-shaped kink structure seen on the western side of the radio core is very unique, and a possible example of plasma instability on such a large scale.
As discussed   in Sect.\ \ref{sec:rmorph} and in the current section, there is a possibility that the eastern side also harbours a kink (twist); however, it is not clearly observed, possibly due to projection effects. 

The radio jets are susceptible to Kelvin--Helmholtz (KH) instabilities when they encounter an ambient medium with contrasting densities and flows \citep{Birkinshaw91,Hardee07,Perucho07}. It has also been shown that   jets having low magnetisation with $\rm Q_{Jet}$ $\sim$\,10$^{45}$ erg~s$^{-1}$ are affected by KH instabilities, which leads to   deceleration and internal turbulence (e.g. \citealt{Dipanjan21p1,Dipanjan21p2}). 
The environment in which the Barbell GRG is found most certainly provides the ideal conditions for the development of KH instabilities, and the observed kink feature could possibly be its manifestation in some form. However, the KH instabilities may not be operating at the scale of the observed kink structure.

Another possibility is that the observed kink feature in the Barbell GRG could also be the site for possible magnetic reconnection \citep{Romanova92}, where the brightening in the jet is observed. This bright or excess emission in the jet can be explained by magnetic reconnection rather than shocks \citep{sironi15} as it has been shown via simulations that while shocks  efficiently dissipate energy, they do not accelerate particles much above the thermal energy.
Recent observation-based studies \citep{shukla2020,Meyer2020}   have also shown compelling evidence of magnetic reconnection in jets of AGNs using high-energy instruments; however, they are on a much smaller scale and are  closer to the AGN.

The current-driven (CD) kink or shear-flow instability can also cause the jets to be unstable, and often leads to brightening in regions of the jets \citep{Chiuderi89,nakamura04}.
Using three-dimensional MHD simulations, \citet{nakamura04,nakamura07} proposed that the main reason for the observed wiggles or kink features in jets is due to CD instability. In this type of instability, a helical structure is developed with a relatively low growth rate. Simulation studies by \citet{Mizuno09,Mizuno12,Mizuno14} have shown that CD instability can stimulate large-scale helical motions in the jets leading to deformation and without complete jet disruption. Here, the distorted magnetic field structure can possibly prompt the creation of magnetic reconnection.
Their simulations also revealed that the CD instability  depends strongly on the magnetic pitch profile and moderately on the density profile. In addition, the velocity shear radius relative to the characteristic radius of the magnetic field determines the static and non-static features of the kink.
In the case of the  Barbell GRG, we observe that the jet on the western side  develops a kink-like feature and  undergoes recollimation before connecting to the WL.
The helical patterns in jets have been observed in a few sources \citep{Pasetto_2021,bruni21,worrall07,Lobanov01} in a different range of spatial scales.

High-resolution studies of radio AGNs using Very Long Baseline Arrays are extremely important for studying jets as they allow us to  closely examine  the region of jet launching.
Using multi-epoch VLBA observations at 15 GHz for BL Lacertae (BL Lac), \citet{cohen15}  show a strong possibility of detecting Alfv\'{e}n waves propagating in the radio jet emanating from the black hole. Their observations  show the existence of a strong transverse component of the magnetic field belonging to the BL Lac jet.
Assuming the magnetic field to be helical, based on previous observational studies \citep{Gabuzda04}, they suggest that the observed features in the BL Lac are due to transverse S (shear) Alfv\'{e}n waves propagating along with the longitudinal component of the magnetic field. These Alfv\'{e}n waves appear to be stimulated by changes in the position angle of the re-collimation shock, which behaves like a whip\footnote{\hyperlink{https://www.nasa.gov/jpl/distant-black-hole-wave-twists-like-giant-whip}{https://www.nasa.gov/jpl/distant-black-hole-wave-twists-like-giant-whip}} in action. Currently, it is unclear whether such disturbances along with jets can propagate to larger scales, but the above study does certainly have some similarities to that of the observed peculiar kink feature in the Barbell GRG.

An interesting mechanism is proposed by \citet{Nakamura01} using helical kink instability to explain the observed wiggles  or kink features in radio jets of AGNs. It is based on the Sweeping Magnetic-Twist model, which conjectures that the AGN core provides the necessary large amount of energy required to produce the systematic magnetic configuration. They  conclude that the   optimum mode of carrying this energy is via the Poynting flux of torsional Alfv\'{e}n wave train (or train for short), which is produced by the interaction of the rotating accretion disk of the AGN and the large-scale magnetic field. The observed wiggles or kink   can then arise from MHD processes under the influence of train;  when it encounters low Alfv\'{e}n velocity, it creates a pinched region and other deformations in the jet. Unlike the models discussed   above, this model explains distortions in jets on a larger scale; as the magnetic twist cannot be created at large distances locally, the train can come along the jet from the radio core of the AGN.

The brightest part of the Barbell GRG is the kink feature, which could possibly be the working surface of supersonic jets in which the bulk kinetic energy is abruptly dissipated within decelerating shock waves. At the location of the shock fronts formed in the jets, the relativistic electrons are efficiently accelerated due to the impact with the ambient surrounding medium. This is evident from the very flat spectral index ($\alpha^{612}_{144}~ = 0.36\pm0.08$ ) of the top (box 1) of the kink feature, as seen on  the righ side of Fig.\ \ref{fig:kinkprspec}, and represents a shock front. The spectral index gradient across the kink with the help of five regions or boxes is shown in the left  panel Fig.\ \ref{fig:kinkprspec}, where it ranges from $\sim$\,0.35 to 0.6. The observed spectral index steepening to the lower downstream side
of shock is consistent with advection and radiative losses. Hence, it clearly depicts the signature of a shock with
in situ particle acceleration at the shock front (the head of the
kink or box 1). \citet{Peacock81}  showed that $\alpha$ of 0.3 to 0.5 can be produced for a wide range of propagation velocities in the case of strong shocks using Fermi acceleration. Using MHD simulations of relativistic jets, \citet{Dipanjan21p1}   show that 
on small scales, KH instabilities affect the dynamics on the jet and on the large scale, the kink modes. They also find that a relatively low-power jet ($\rm Q_{Jet}$ $\sim$\,10$^{44}$ erg~s$^{-1}$) with strong magnetisation is not stable against instabilities occurring due to kink mode and hence leads to strong bending of the jet head. This in turn leads to the formation of peculiar shock structures at the jet head, which is very similar to the  Barbell GRG kink feature.
However, some simulation studies have shown that jets with relativistic plasma that are magnetically dominated are not efficient sites for shock acceleration \citep{Sironi13,sironi15}. 

Although we  considered a number of possibilities suggested in the literature for the formation of a kink in the flow of a jet, one also needs to consider the overall structure of the source. The presence of a kink and a similar structure on the opposite side (which we  call a twist) is reminiscent of precessing jets, although there could be instabilities and shocks in the flow of the jets.
\citet{GowerModel82} presented a simple yet effective kinematical model of precessing twin jets with relativistic bulk velocity that explains peculiar radio jets with twists and turns of several radio galaxies and quasars. Parts of the Barbell GRG resemble the radio galaxy 3C~449, whose peculiar jet morphology with mirror symmetry was explained by \citet{Gower3c44982} with a simple precessing beam model having relativistic bulk motion. This remains the most promising possibility for the overall structure of the source. 

If the observed asymmetry of the jets is due to relativistic beaming, as appears to be the case for almost all radio jets \citep[for a review, see][]{Saikia22}, the western jet is approaching us while the eastern jet is receding. For a collinear symmetric source, the twist which is on the receding side should appear nearer from the core than the kink which is on the approaching side. This is consistent with the observations.

It is also interesting to note that the EL, which is on the receding side, has a flux density that is on average a factor of $\sim$\,1.7 higher than that of the WL. The brightness of the EL and WL  estimated from the LOFAR 144 MHz image are similar, the values being 3.48 and 3.68 mJy/arcsec$^2$, respectively, which are within $\sim$5\%. As the EL is larger, this leads to a higher value of flux density and hence luminosity. A higher magnetic field in the EL compared with the WL would lead to increased synchrotron luminosity in the EL. However,  if the magnetic fields are similar to the equipartition values, this is an unlikely explanation as the EL has a marginally lower value than the WL (see Table~\ref{Tab:Specage}). Reacceleration of particles as the lobes interact with the intracluster medium could lead to increased radio emission. The EL has a slightly flatter integrated spectral index between 144 and 1400 MHz, as  expected in such a scenario, and it would be interesting to explore this possibility further using data at more frequencies. Another possibility is that in addition to the effects of relativistic motion, the kink on the western side could be brighter due to greater dissipation of jet energy in the kink, leading to a decreased energy supply to the WL. 

The precessing jet model is the most promising possibility for the overall structure of the Barbell GRG. As discussed earlier, precessing jet models are applicable to a wide variety of sources belonging to both FRI and FRII sources. \citet{GowerModel82} and \citet{Gower3c44982} have made a detailed model for 3C449 and also given parameters for 3C31. Although there are similarities, there are significant differences between these two galaxies and the Barbell GRG. Galaxies 3C449 and 3C31 are FRI galaxies where the jets expand to form diffuse plumes of emission, while the Barbell GRG has collimated jets, as in FRII galaxies, which lead towards diffuse lobes. For a more detailed modelling of the Barbell GRG, which is beyond the scope of the present paper, deep X-ray observations to determine the gas distribution, spectroscopic observations to determine the kinematics of the galaxy cluster, high-resolution radio images of the jets, and radio polarization information of the jets and the lobes, all of which we do not have at present, would be very valuable.

\begin{figure*}
\centering
\includegraphics[scale=0.372]{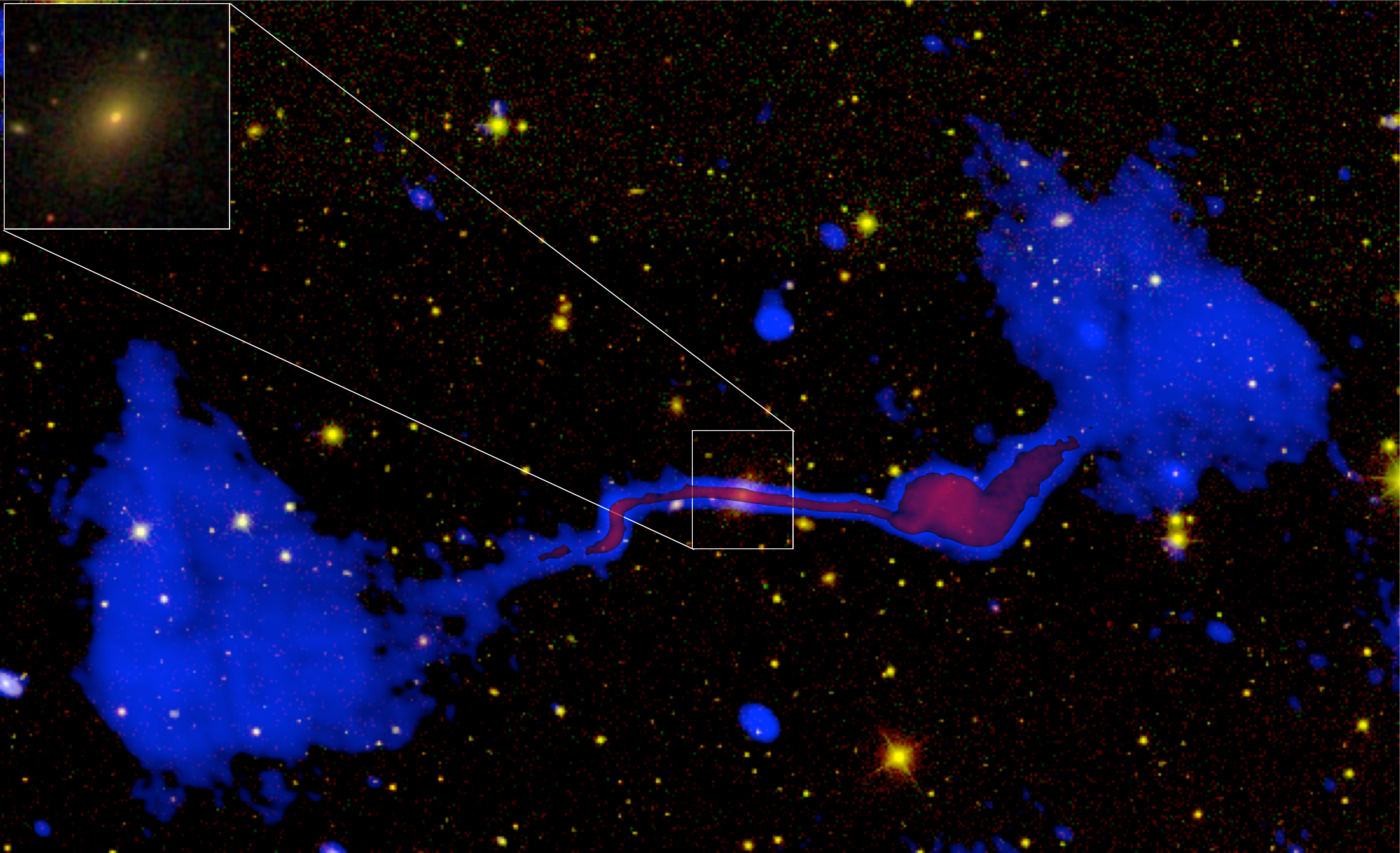}
\caption{Optical and radio colour composite image. The background yellow image represents the SDSS, the blue and red represents radio emission as seen in GMRT 323 MHz  and 610 MHz (high resolution) maps, respectively. At the top left a zoomed-in view of the host galaxy (BCG SDSSJ223301.30+131502.5) is shown.}
\label{fig:composite}
\end{figure*}

%%%%%%%%%%%%%%%%%%%%%%%%%%%%%%%%%%%%%%%%%%%%%%%%%%%%%%%%%%%%%%%%%%%%%%%%%%
\begin{table*}
\begin{center}
\setlength{\tabcolsep}{15pt}

% \begin{minipage}{130mm}
\captionsetup{width=18cm}
\caption{WHLJ223301.3+131503 galaxy cluster parameters obtained from \citet{whl12}: (1) $\rm R_{200}$: virial radius of the galaxy cluster; (2) $\rm R_{L*}$: richness of the galaxy cluster,   defined as $\rm R_{L*}$ = $\rm L_{200}$/L$_*$, where L$_*$ is the characteristic luminosity of galaxies in the r band and $\rm L_{200}$ the total r-band luminosity within the radius of $\rm R_{200}$; (3)$\rm N_{200}$: number of galaxies observed within $\rm R_{200}$; (4) $\rm M_{200}$: mass within $\rm R_{200}$; (5) Volume: volume of cluster within $\rm R_{200}$; (6) Density: within $\rm R_{200}$;  (7) L$_{\rm X}$: cluster X-ray luminosity; and (8)  T$_{\rm X}$: temperature in the X-ray band, 0.1 to 2.4 keV. }
\setlength{\extrarowheight}{3pt}
\begin{tabular}{lccccccc}
%\toprule[1.5pt]
\hline
 $\rm R_{200}$ & $\rm R_{L*}$ & $\rm N_{200}$ & $\rm M_{200}$ & V & Density & $\rm L_X$ & $\rm T_X$ \\  
  (Mpc) & &  & ($\rm M_{\odot}$) & ($\rm Mpc^{3}$) &  ($\rm M_{\odot}$ / $\rm Mpc^{3}$ ) & (erg s$^{-1}$ ) & (keV)\\  
  (1) & (2) & (3) & (4) & (5) & (6) & (7) & (8) \\ \hline 
1.02 & 20.04 & 13 & 1.1 $\times$ $10^{14}$ & 4.4 & 2.4 $\times$ $\rm 10^{13}$ & 3 $\times$ $\rm 10^{43}$ & 2.3\\
\hline
    
\end{tabular}
\label{Tab:env}
% \end{minipage}
\end{center}
\end{table*}
%%%%%%%%%%%%%%%%%%%%%%%%%%%%%%%%%%%%%%%%%%%%%%%%%%%%%%%%%%%%%%%%%%%%%%%%%%

\subsection{Environment of the Barbell GRG}
It is essential to study the environment of GRGs as one of the most favoured explanations for their gigantic sizes is   an underdense environment \citep{mack98}. Another important reason to study their environment is to understand its effect on their morphologies. Some GRG studies \citep{Schoenmakers_spec,lara2001,ravi08,Pirya} have indicated that the observed asymmetries in the morphologies of the GRGs trace the asymmetries in their environment. Hence, GRGs, thanks to their large sizes, become a natural probe of environmental asymmetries over megaparsec scales \citep{ravi08,Safouris09}.

Using data from the literature, \citet{Komberg09} presented observational evidence of several GRGs residing in a group or cluster environment.
Recently, large GRG samples from the LoTSS and NVSS have revealed nearly 60 such examples of GRGs residing at the centres (with BCGs as their host galaxies) of galaxy clusters \citep{PDLOTSS,sagan1} and 13 from the Rapid ASKAP Continuum Survey  \citep{Andernach2021}. GRGs hosted by BCGs in cluster environments are relatively uncommon when considering the total known GRG population, and as shown by \citet{sagan1}, only $\sim$\,15\% of the known GRG population are associated with BCGs.

As briefly mentioned in Sect.\ \ref{sec:intro}, the Barbell GRG resides at the centre of WHLJ223301.3+131503 galaxy cluster. This galaxy cluster was identified by \citet{whl12}, who   also provided the additional information about the galaxy cluster  presented in Table\ \ref{Tab:env}. \citet{whl12} cross-matched their galaxy cluster catalogue with the ROSAT X-ray survey and X-ray cluster database \citep[BAX;][]{Sadat2004} to find L$_{\rm X}$ and T$_{\rm X}$ for the galaxy clusters detected in the combined sample of ROSAT and BAX, and then used these data to calibrate the correlation between $\rm R_{L*}$ and L$_{\rm X}$. They then used this correlation (see details in Sect. 3.6 of \citealt{whl12})   to estimate L$_{\rm X}$  and T$_{\rm X}$  for the galaxy clusters  not directly detected by ROSAT, including WHLJ223301.3+131503 (Fig.\ \ref{fig:composite}) in which the Barbell GRG resides.

The mass ($\rm M_{200}$ $\sim$\,1.1 $\times$ 10$^{14}$ $\rm M_{\odot}$) of the galaxy cluster WHLJ223301.3+131503 reflects that it is either a low-mass galaxy cluster or a massive galaxy group. 
The non-detection of the cluster in the ROSAT sky survey is consistent with this.
Therefore, in the case of the Barbell GRG, it appears,  with the right combination of kinetic jet power and  ambient medium, that the collimated jets travel outwards to form a megaparsec-scale radio galaxy. The value of the virial radius $\rm R_{200}$ of $\sim$\,1.02 Mpc for this cluster \citep{whl12} suggests that  GRG-J2233+1315 lies within this radius. The cluster environment is likely to affect the structure of the Barbell GRG. However, as mentioned earlier, deep X-ray observations to image the cluster gas distribution and its properties are required to explore this in detail.

%%%%%%%%%%%%%%%%%%%%%% Section Change %%%%%%%%%%%%%%%%%%%%%%%%%%%%%%%%%%%%%%%
\section{Summary}
We have presented our study on a 1.83 Mpc large GRG called the Barbell GRG, which is hosted in a relatively dense environment of a galaxy cluster.
Our optical spectroscopic observations of the host galaxy have established the redshift (0.09956) of the GRG, and hence its projected linear size. Our analysis also indicates a low SFR ($\sim$\,$10^{-3} \, \rm M_{\odot} yr^{-1}$) for the host galaxy, consistent with its BCG nature.
Our deep and high-resolution radio images from GMRT and LOFAR reveal a $\sim$\,237 kpc jet emanating from the radio core and leading to $\sim$\,100\,kpc kink.
The multi-frequency radio data enabled us to estimate the magnetic field and spectral ages of the diffuse lobes.
The classical and revised equivalent magnetic field for the WL are $\sim$\,3.7 $\rm \muup$\,G and $\sim$\, 6.8\,$\rm \muup$\,G, respectively. For the EL the values are  $\sim$\,3.2 $\rm \muup$\,G (classical) and $\sim$\, 5.7\,$\rm \muup$\,G (revised). The corresponding classical and revised spectral ages for the WL are respectively  $\sim$\,183 \,Myr and $\sim$\,115 \,Myr, and for the EL $\sim$\,193 \,Myr and $\sim$\,135. The discovery of the $\sim$\,100\,kpc kink structure from our study provides a unique opportunity for testing various MHD models on large scales. We have discussed a few possibilities, such as the development of instabilities, magnetic reconnection, and precession of jet axis, which could have led to the creation of these structures.
Precessing jets in an active galaxy residing in a cluster environment, with shocks and instabilities in the flow of the jets, is a promising possibility to understand the overall misaligned structure of the source. Further observations at X-ray wavelengths to study the gas distribution, spectroscopic data to understand the cluster dynamics, high-resolution radio observations of the jets, and detailed polarization observations would be useful for a detailed modelling and understanding of the source. 

%%%%%%%%%%%%%%%%%%%%%%% Acknowledgements %%%%%%%%%%%%%%%%%%%%%%%%%%%%%%%%%%%%%
\section*{Acknowledgements}
We thank an anonymous referee for her/his valuable comments which have helped to significantly improve the manuscript. We also thank Aayush Saxena, Shishir Sankhyayan, Sumana Nandi, and Mousumi Mahato for their help. PD, JB, and FC gratefully acknowledge generous support from the Indo-French Centre for the Promotion of Advanced Research 
(Centre Franco-Indien pour la Promotion de la Recherche Avan\'{c}ee) under programme no. 5204-2 (2015-2018). We thank IUCAA 
(especially Radio Physics Lab\footnote{\url{http://www.iucaa.in/~rpl/}}), Pune for providing all the facilities during the period the work was carried out. 
We gratefully acknowledge the use of Edward (Ned) Wright's online Cosmology Calculator. 

We thank the staff of the GMRT that made these observations possible. GMRT is run by the National Centre for Radio Astrophysics of the Tata Institute of Fundamental Research.
LOFAR \citep{lofar} is the Low-Frequency Array designed and constructed by ASTRON. It has observing, data processing, and data storage facilities in several countries, which are owned by various parties (each with their own funding sources), and that are collectively operated by the ILT foundation under a joint scientific policy. The ILT resources have benefited from the following recent major funding sources: CNRS-INSU, Observatoire de Paris and Universit\'{e} d'Orl\'{e}ans, France; BMBF, MIWF-NRW, MPG, Germany; Science Foundation Ireland (SFI), Department of Business, Enterprise and Innovation (DBEI), Ireland; NWO, The Netherlands; The Science and Technology Facilities Council, UK; Ministry of Science and Higher Education, Poland; The Istituto Nazionale di Astrofisica (INAF), Italy. This research made use of the LOFAR-UK computing facility located at the University of Hertfordshire and supported by STFC [ST/P000096/1], and of the LOFAR-IT computing infrastructure supported and operated by INAF, and by the Physics Dept. of Turin University (under the agreement with Consorzio Interuniversitario per la Fisica Spaziale) at the C3S Supercomputing Centre, Italy. This article is based on observations made in the Observatorios de Canarias del IAC with the WHT operated on the island of La Palma by the Isaac Newton Group of Telescopes (ING) in the Observatorio del Roque de los Muchachos.

We acknowledge that this work has made use of  \textsc{aplpy} \citep{apl}.

%%%%%%%%%%%%%%%%%%%% REFERENCES %%%%%%%%%%%%%%%%%%
\bibliographystyle{aa} 
\bibliography{Barbell_Dabhade.bib}

\begin{thebibliography}{159}
\expandafter\ifx\csname natexlab\endcsname\relax\def\natexlab#1{#1}\fi

\bibitem[{{Abolfathi} {et~al.}(2018){Abolfathi}, {Aguado}, {Aguilar}, {Allende
  Prieto}, {Almeida}, {Ananna}, {Anders}, {Anderson}, {Andrews}, {Anguiano},
  {Arag{\'o}n-Salamanca}, {Argudo-Fern{\'a}ndez}, {Armengaud}, {Ata},
  {Aubourg}, {Avila-Reese}, {Badenes}, {Bailey}, {Balland}, {Barger},
  {Barrera-Ballesteros}, {Bartosz}, {Bastien}, {Bates}, {Baumgarten},
  {Bautista}, {Beaton}, {Beers}, {Belfiore}, {Bender}, {Bernardi}, {Bershady},
  {Beutler}, {Bird}, {Bizyaev}, {Blanc}, {Blanton}, {Blomqvist}, {Bolton},
  {Boquien}, {Borissova}, {Bovy}, {Andres Bradna Diaz}, {Brandt}, {Brinkmann},
  {Brownstein}, {Bundy}, {Burgasser}, {Burtin}, {Busca}, {Ca{\~n}as},
  {Cano-D{\'\i}az}, {Cappellari}, {Carrera}, {Casey}, {Cervantes Sodi}, {Chen},
  {Cherinka}, {Chiappini}, {Doohyun Choi}, {Chojnowski}, {Chuang}, {Chung},
  {Clerc}, {Cohen}, {Comerford}, {Comparat}, {Correa do Nascimento}, {da
  Costa}, {Cousinou}, {Covey}, {Crane}, {Cruz-Gonzalez}, {Cunha}, {da Silva
  Ilha}, {Damke}, {Darling}, {Davidson}, {Dawson}, {de Icaza Lizaola}, {de la
  Macorra}, {de la Torre}, {De Lee}, {de Sainte Agathe}, {Deconto Machado},
  {Dell'Agli}, {Delubac}, {Diamond-Stanic}, {Donor}, {Downes}, {Drory}, {du Mas
  des Bourboux}, {Duckworth}, {Dwelly}, {Dyer}, {Ebelke}, {Davis Eigenbrot},
  {Eisenstein}, {Elsworth}, {Emsellem}, {Eracleous}, {Erfanianfar},
  {Escoffier}, {Fan}, {Fern{\'a}ndez Alvar}, {Fernandez-Trincado}, {Fernand o
  Cirolini}, {Feuillet}, {Finoguenov}, {Fleming}, {Font-Ribera}, {Freischlad},
  {Frinchaboy}, {Fu}, {G{\'o}mez Maqueo Chew}, {Galbany}, {Garc{\'\i}a
  P{\'e}rez}, {Garcia-Dias}, {Garc{\'\i}a-Hern{\'a}ndez}, {Garma Oehmichen},
  {Gaulme}, {Gelfand }, {Gil-Mar{\'\i}n}, {Gillespie}, {Goddard}, {Gonz{\'a}lez
  Hern{\'a}ndez}, {Gonzalez-Perez}, {Grabowski}, {Green}, {Grier}, {Gueguen},
  {Guo}, {Guy}, {Hagen}, {Hall}, {Harding}, {Hasselquist}, {Hawley}, {Hayes},
  {Hearty}, {Hekker}, {Hernand ez}, {Hernandez Toledo}, {Hogg},
  {Holley-Bockelmann}, {Holtzman}, {Hou}, {Hsieh}, {Hunt}, {Hutchinson},
  {Hwang}, {Jimenez Angel}, {Johnson}, {Jones}, {J{\"o}nsson}, {Jullo}, {Khan},
  {Kinemuchi}, {Kirkby}, {Kirkpatrick}, {Kitaura}, {Knapp}, {Kneib},
  {Kollmeier}, {Lacerna}, {Lane}, {Lang}, {Law}, {Le Goff}, {Lee}, {Li}, {Li},
  {Lian}, {Liang}, {Lima}, {Lin}, {Long}, {Lucatello}, {Lundgren}, {Mackereth},
  {MacLeod}, {Mahadevan}, {Maia}, {Majewski}, {Manchado}, {Maraston},
  {Mariappan}, {Marques-Chaves}, {Masseron}, {Masters}, {McDermid}, {McGreer},
  {Melendez}, {Meneses-Goytia}, {Merloni}, {Merrifield}, {Meszaros}, {Meza},
  {Minchev}, {Minniti}, {Mueller}, {Muller-Sanchez}, {Muna}, {Mu{\~n}oz},
  {Myers}, {Nair}, {Nand ra}, {Ness}, {Newman}, {Nichol}, {Nidever},
  {Nitschelm}, {Noterdaeme}, {O'Connell}, {Oelkers}, {Oravetz}, {Oravetz},
  {Ort{\'\i}z}, {Osorio}, {Pace}, {Padilla}, {Palanque-Delabrouille},
  {Palicio}, {Pan}, {Pan}, {Parikh}, {P{\^a}ris}, {Park}, {Peirani},
  {Pellejero-Ibanez}, {Penny}, {Percival}, {Perez-Fournon}, {Petitjean},
  {Pieri}, {Pinsonneault}, {Pisani}, {Prada}, {Prakash}, {Queiroz}, {Raddick},
  {Raichoor}, {Barboza Rembold}, {Richstein}, {Riffel}, {Riffel}, {Rix},
  {Robin}, {Rodr{\'\i}guez Torres}, {Rom{\'a}n-Z{\'u}{\~n}iga}, {Ross},
  {Rossi}, {Ruan}, {Ruggeri}, {Ruiz}, {Salvato}, {S{\'a}nchez}, {S{\'a}nchez},
  {Sanchez Almeida}, {S{\'a}nchez-Gallego}, {Santana Rojas}, {Santiago},
  {Schiavon}, {Schimoia}, {Schlafly}, {Schlegel}, {Schneider}, {Schuster},
  {Schwope}, {Seo}, {Serenelli}, {Shen}, {Shen}, {Shetrone}, {Shull}, {Silva
  Aguirre}, {Simon}, {Skrutskie}, {Slosar}, {Smethurst}, {Smith}, {Sobeck},
  {Somers}, {Souter}, {Souto}, {Spindler}, {Stark}, {Stassun}, {Steinmetz},
  {Stello}, {Storchi-Bergmann}, {Streblyanska}, {Stringfellow}, {Su{\'a}rez},
  {Sun}, {Szigeti}, {Taghizadeh-Popp}, {Talbot}, {Tang}, {Tao}, {Tayar},
  {Tembe}, {Teske}, {Thakar}, {Thomas}, {Tissera}, {Tojeiro}, {Tremonti},
  {Troup}, {Urry}, {Valenzuela}, {van den Bosch}, {Vargas-Gonz{\'a}lez},
  {Vargas-Maga{\~n}a}, {Vazquez}, {Villanova}, {Vogt}, {Wake}, {Wang},
  {Weaver}, {Weijmans}, {Weinberg}, {Westfall}, {Whelan}, {Wilcots}, {Wild},
  {Williams}, {Wilson}, {Wood-Vasey}, {Wylezalek}, {Xiao}, {Yan}, {Yang},
  {Ybarra}, {Y{\`e}che}, {Zakamska}, {Zamora}, {Zarrouk}, {Zasowski}, {Zhang},
  {Zhao}, {Zhao}, {Zheng}, {Zheng}, {Zhou}, {Zhu}, {Zinn}, \& {Zou}}]{sdssdr14}
{Abolfathi}, B., {Aguado}, D.~S., {Aguilar}, G., {et~al.} 2018, The
  Astrophysical Journal Supplement Series, 235, 42

\bibitem[{{Andernach} {et~al.}(2021){Andernach}, {Jim{\'e}nez-Andrade}, \&
  {Willis}}]{Andernach2021}
{Andernach}, H., {Jim{\'e}nez-Andrade}, E.~F., \& {Willis}, A.~G. 2021,
  Galaxies, 9, 99

\bibitem[{{Anders} \& {Grevesse}(1989)}]{Anders89}
{Anders}, E. \& {Grevesse}, N. 1989, \gca, 53, 197

\bibitem[{{Bagchi} {et~al.}(2007){Bagchi}, {Gopal-Krishna}, {Krause}, \&
  {Joshi}}]{bagchi07}
{Bagchi}, J., {Gopal-Krishna}, {Krause}, M., \& {Joshi}, S. 2007, \apjl, 670,
  L85

\bibitem[{{Baum} {et~al.}(1997){Baum}, {O'Dea}, {Giovannini}, {Cotton}, {de
  Koff}, {Feretti}, {Golombek}, {Lara}, {Macchetto}, {Miley}, {Sparks},
  {Venturi}, \& {Komissarov}}]{baum97}
{Baum}, S.~A., {O'Dea}, C.~P., {Giovannini}, G., {et~al.} 1997, \apj, 483, 178

\bibitem[{{Beck} \& {Krause}(2005)}]{Beck_Krause}
{Beck}, R. \& {Krause}, M. 2005, Astronomische Nachrichten, 326, 414

\bibitem[{{Beskin} {et~al.}(1998){Beskin}, {Kuznetsova}, \&
  {Rafikov}}]{Beskin1998}
{Beskin}, V.~S., {Kuznetsova}, I.~V., \& {Rafikov}, R.~R. 1998, \mnras, 299,
  341

\bibitem[{{Best} \& {Heckman}(2012)}]{bh12rgs}
{Best}, P.~N. \& {Heckman}, T.~M. 2012, \mnras, 421, 1569

\bibitem[{{Biju} {et~al.}(2017){Biju}, {Bagchi}, {Ishwara-Chandra},
  {Pandey-Pommier}, {Jacob}, {Patil}, {Kumar}, {Pandge}, {Dabhade}, {Gaikwad},
  {Dhurde}, {Abraham}, {Vivek}, {Mahabal}, \& {Djorgovski}}]{bijua407}
{Biju}, K.~G., {Bagchi}, J., {Ishwara-Chandra}, C.~H., {et~al.} 2017, \mnras,
  471, 617

\bibitem[{{Biju} {et~al.}(2014){Biju}, {Pandey-Pommier}, {Sunilkumar},
  {Dhurde}, {Bagchi}, {Ishwara-Chandra}, \& {Jacob}}]{Biju14}
{Biju}, K.~G., {Pandey-Pommier}, M., {Sunilkumar}, P., {et~al.} 2014, in
  Astronomical Society of India Conference Series, Vol.~13, Astronomical
  Society of India Conference Series, 155--156

\bibitem[{{Birkinshaw}(1991)}]{Birkinshaw91}
{Birkinshaw}, M. 1991, \mnras, 252, 505

\bibitem[{{B{\^\i}rzan} {et~al.}(2008){B{\^\i}rzan}, {McNamara}, {Nulsen},
  {Carilli}, \& {Wise}}]{birzan08}
{B{\^\i}rzan}, L., {McNamara}, B.~R., {Nulsen}, P.~E.~J., {Carilli}, C.~L., \&
  {Wise}, M.~W. 2008, \apj, 686, 859

\bibitem[{{B{\^\i}rzan} {et~al.}(2004){B{\^\i}rzan}, {Rafferty}, {McNamara},
  {Wise}, \& {Nulsen}}]{birzan04}
{B{\^\i}rzan}, L., {Rafferty}, D.~A., {McNamara}, B.~R., {Wise}, M.~W., \&
  {Nulsen}, P.~E.~J. 2004, \apj, 607, 800

\bibitem[{{Blandford} {et~al.}(2019){Blandford}, {Meier}, \&
  {Readhead}}]{Blandford2019}
{Blandford}, R., {Meier}, D., \& {Readhead}, A. 2019, \araa, 57, 467

\bibitem[{{Blandford} \& {Rees}(1974)}]{Blanford_Rees74}
{Blandford}, R.~D. \& {Rees}, M.~J. 1974, \mnras, 169, 395

\bibitem[{{Br{\"u}ggen} \& {Kaiser}(2002)}]{brugenanture02}
{Br{\"u}ggen}, M. \& {Kaiser}, C.~R. 2002, \nat, 418, 301

\bibitem[{{Brunetti} {et~al.}(1997){Brunetti}, {Setti}, \&
  {Comastri}}]{Brunetti1997}
{Brunetti}, G., {Setti}, G., \& {Comastri}, A. 1997, \aap, 325, 898

\bibitem[{{Bruni} {et~al.}(2021){Bruni}, {G{\'o}mez}, {Vega-Garc{\'\i}a},
  {Lobanov}, {Fuentes}, {Savolainen}, {Kovalev}, {Perucho}, {Mart{\'\i}},
  {Anderson}, {Edwards}, {Gurvits}, {Lisakov}, {Pushkarev}, {Sokolovsky}, \&
  {Zensus}}]{bruni21}
{Bruni}, G., {G{\'o}mez}, J.~L., {Vega-Garc{\'\i}a}, L., {et~al.} 2021, \aap,
  654, A27

\bibitem[{{Bruni} {et~al.}(2020){Bruni}, {Panessa}, {Bassani}, {Dallacasa},
  {Venturi}, {Saripalli}, {Brienza}, {Hern{\'a}ndez-Garc{\'\i}a}, {Chiaraluce},
  {Ursini}, {Bazzano}, {Malizia}, \& {Ubertini}}]{bruni20}
{Bruni}, G., {Panessa}, F., {Bassani}, L., {et~al.} 2020, \mnras, 494, 902

\bibitem[{{Bruzual} \& {Charlot}(2003)}]{bruzal2003}
{Bruzual}, G. \& {Charlot}, S. 2003, \mnras, 344, 1000

\bibitem[{{Buttiglione} {et~al.}(2010){Buttiglione}, {Capetti}, {Celotti},
  {Axon}, {Chiaberge}, {Macchetto}, \& {Sparks}}]{Buttiglione2010}
{Buttiglione}, S., {Capetti}, A., {Celotti}, A., {et~al.} 2010, \aap, 509, A6

\bibitem[{{Byler} {et~al.}(2017){Byler}, {Dalcanton}, {Conroy}, \&
  {Johnson}}]{Byler17}
{Byler}, N., {Dalcanton}, J.~J., {Conroy}, C., \& {Johnson}, B.~D. 2017, \apj,
  840, 44

\bibitem[{{Calzetti} {et~al.}(2000){Calzetti}, {Armus}, {Bohlin}, {Kinney},
  {Koornneef}, \& {Storchi-Bergmann}}]{Calzetti00}
{Calzetti}, D., {Armus}, L., {Bohlin}, R.~C., {et~al.} 2000, \apj, 533, 682

\bibitem[{{Cantwell} {et~al.}(2020){Cantwell}, {Bray}, {Croston}, {Scaife},
  {Mulcahy}, {Best}, {Br{\"u}ggen}, {Brunetti}, {Callingham}, {Clarke},
  {Hardcastle}, {Harwood}, {Heald}, {Heesen}, {Iacobelli}, {Jamrozy},
  {Morganti}, {Orr{\'u}}, {O'Sullivan}, {Riseley}, {R{\"o}ttgering},
  {Shulevski}, {Sridhar}, {Tasse}, \& {Van Eck}}]{cantwell20}
{Cantwell}, T.~M., {Bray}, J.~D., {Croston}, J.~H., {et~al.} 2020, \mnras, 495,
  143

\bibitem[{{Carnall} {et~al.}(2018){Carnall}, {McLure}, {Dunlop}, \&
  {Dav{\'e}}}]{Carnall18}
{Carnall}, A.~C., {McLure}, R.~J., {Dunlop}, J.~S., \& {Dav{\'e}}, R. 2018,
  \mnras, 480, 4379

\bibitem[{{Chiuderi} {et~al.}(1989){Chiuderi}, {Pietrini}, \&
  {Ciamponi}}]{Chiuderi89}
{Chiuderi}, C., {Pietrini}, P., \& {Ciamponi}, G.~T. 1989, \apj, 339, 70

\bibitem[{{Cohen} {et~al.}(2015){Cohen}, {Meier}, {Arshakian}, {Clausen-Brown},
  {Homan}, {Hovatta}, {Kovalev}, {Lister}, {Pushkarev}, {Richards}, \&
  {Savolainen}}]{cohen15}
{Cohen}, M.~H., {Meier}, D.~L., {Arshakian}, T.~G., {et~al.} 2015, \apj, 803, 3

\bibitem[{{Condon} {et~al.}(1998){Condon}, {Cotton}, {Greisen}, {Yin},
  {Perley}, {Taylor}, \& {Broderick}}]{nvss}
{Condon}, J.~J., {Cotton}, W.~D., {Greisen}, E.~W., {et~al.} 1998, \aj, 115,
  1693

\bibitem[{{Dabhade} {et~al.}(2020{\natexlab{a}}){Dabhade}, {Combes},
  {Salom{\'e}}, {Bagchi}, \& {Mahato}}]{sagan2}
{Dabhade}, P., {Combes}, F., {Salom{\'e}}, P., {Bagchi}, J., \& {Mahato}, M.
  2020{\natexlab{a}}, \aap, 643, A111

\bibitem[{{Dabhade} {et~al.}(2017){Dabhade}, {Gaikwad}, {Bagchi},
  {Pandey-Pommier}, {Sankhyayan}, \& {Raychaudhury}}]{D17}
{Dabhade}, P., {Gaikwad}, M., {Bagchi}, J., {et~al.} 2017, \mnras, 469, 2886

\bibitem[{{Dabhade} {et~al.}(2020{\natexlab{b}}){Dabhade}, {Mahato}, {Bagchi},
  {Saikia}, {Combes}, {Sankhyayan}, {R{\"o}ttgering}, {Ho}, {Gaikwad},
  {Raychaudhury}, {Vaidya}, \& {Guiderdoni}}]{sagan1}
{Dabhade}, P., {Mahato}, M., {Bagchi}, J., {et~al.} 2020{\natexlab{b}}, \aap,
  642, A153

\bibitem[{{Dabhade} {et~al.}(2020{\natexlab{c}}){Dabhade}, {R{\"o}ttgering},
  {Bagchi}, {Shimwell}, {Hardcastle}, {Sankhyayan}, {Morganti}, {Jamrozy},
  {Shulevski}, \& {Duncan}}]{PDLOTSS}
{Dabhade}, P., {R{\"o}ttgering}, H.~J.~A., {Bagchi}, J., {et~al.}
  2020{\natexlab{c}}, \aap, 635, A5

\bibitem[{{Dabhade} {et~al.}(2022){Dabhade}, {Saikia}, \& {Mahato}}]{DSM22}
{Dabhade}, P., {Saikia}, D.~J., \& {Mahato}, M. 2022, arXiv e-prints, SKA-India
  special issue in Journal of Astrophysics and Astronomy-In press,
  arXiv:2208.02130

\bibitem[{{de Gasperin} {et~al.}(2019){de Gasperin}, {Dijkema}, {Drabent},
  {Mevius}, {Rafferty}, {van Weeren}, {Br{\"u}ggen}, {Callingham}, {Emig},
  {Heald}, {Intema}, {Morabito}, {Offringa}, {Oonk}, {Orr{\`u}},
  {R{\"o}ttgering}, {Sabater}, {Shimwell}, {Shulevski}, \&
  {Williams}}]{deGasperin2019}
{de Gasperin}, F., {Dijkema}, T.~J., {Drabent}, A., {et~al.} 2019, \aap, 622,
  A5

\bibitem[{{Delhaize} {et~al.}(2021){Delhaize}, {Heywood}, {Prescott}, {Jarvis},
  {Delvecchio}, {Whittam}, {White}, {Hardcastle}, {Hale}, {Afonso}, {Ao},
  {Brienza}, {Br{\"u}ggen}, {Collier}, {Daddi}, {Glowacki}, {Maddox},
  {Morabito}, {Prandoni}, {Randriamanakoto}, {Sekhar}, {An}, {Adams}, {Blyth},
  {Bowler}, {Leeuw}, {Marchetti}, {Randriamampandry}, {Thorat}, {Seymour},
  {Smirnov}, {Taylor}, {Tasse}, \& {Vaccari}}]{Delhaize21}
{Delhaize}, J., {Heywood}, I., {Prescott}, M., {et~al.} 2021, \mnras, 501, 3833

\bibitem[{{Dopita} {et~al.}(2000){Dopita}, {Kewley}, {Heisler}, \&
  {Sutherland}}]{Dopita2000}
{Dopita}, M.~A., {Kewley}, L.~J., {Heisler}, C.~A., \& {Sutherland}, R.~S.
  2000, \apj, 542, 224

\bibitem[{{Evans} {et~al.}(2005){Evans}, {Hardcastle}, {Croston}, {Worrall}, \&
  {Birkinshaw}}]{Evans05}
{Evans}, D.~A., {Hardcastle}, M.~J., {Croston}, J.~H., {Worrall}, D.~M., \&
  {Birkinshaw}, M. 2005, \mnras, 359, 363

\bibitem[{{Fabian} {et~al.}(2002){Fabian}, {Celotti}, {Blundell}, {Kassim}, \&
  {Perley}}]{fabian02}
{Fabian}, A.~C., {Celotti}, A., {Blundell}, K.~M., {Kassim}, N.~E., \&
  {Perley}, R.~A. 2002, \mnras, 331, 369

\bibitem[{{Fanaroff} \& {Riley}(1974)}]{FR74}
{Fanaroff}, B.~L. \& {Riley}, J.~M. 1974, \mnras, 167, 31P

\bibitem[{{Feretti} {et~al.}(1999){Feretti}, {Perley}, {Giovannini}, \&
  {Andernach}}]{Feretti99}
{Feretti}, L., {Perley}, R., {Giovannini}, G., \& {Andernach}, H. 1999, \aap,
  341, 29

\bibitem[{{Ferland} {et~al.}(2017){Ferland}, {Chatzikos}, {Guzm{\'a}n},
  {Lykins}, {van Hoof}, {Williams}, {Abel}, {Badnell}, {Keenan}, {Porter}, \&
  {Stancil}}]{Ferland17}
{Ferland}, G.~J., {Chatzikos}, M., {Guzm{\'a}n}, F., {et~al.} 2017, \rmxaa, 53,
  385

\bibitem[{{Gabuzda} {et~al.}(2004){Gabuzda}, {Murray}, \& {Cronin}}]{Gabuzda04}
{Gabuzda}, D.~C., {Murray}, {\'E}., \& {Cronin}, P. 2004, \mnras, 351, L89

\bibitem[{{Giacintucci} {et~al.}(2021){Giacintucci}, {Clarke}, {Kassim},
  {Peters}, \& {Polisensky}}]{Giacintucci2021}
{Giacintucci}, S., {Clarke}, T., {Kassim}, N.~E., {Peters}, W., \&
  {Polisensky}, E. 2021, Galaxies, 9, 108

\bibitem[{{Godfrey} \& {Shabala}(2013)}]{Godfrey13}
{Godfrey}, L.~E.~H. \& {Shabala}, S.~S. 2013, \apj, 767, 12

\bibitem[{{Gopal-Krishna} {et~al.}(1989){Gopal-Krishna}, {Wiita}, \&
  {Saripalli}}]{gk89}
{Gopal-Krishna}, {Wiita}, P.~J., \& {Saripalli}, L. 1989, \mnras, 239, 173

\bibitem[{{Govoni} \& {Feretti}(2004)}]{Govoni04}
{Govoni}, F. \& {Feretti}, L. 2004, International Journal of Modern Physics D,
  13, 1549

\bibitem[{{Gower} {et~al.}(1982){Gower}, {Gregory}, {Unruh}, \&
  {Hutchings}}]{GowerModel82}
{Gower}, A.~C., {Gregory}, P.~C., {Unruh}, W.~G., \& {Hutchings}, J.~B. 1982,
  \apj, 262, 478

\bibitem[{{Gower} \& {Hutchings}(1982)}]{Gower3c44982}
{Gower}, A.~C. \& {Hutchings}, J.~B. 1982, \apjl, 258, L63

\bibitem[{{Greisen}(2003)}]{Greisen03}
{Greisen}, E.~W. 2003, {AIPS, the VLA, and the VLBA}, Vol. 285, 109

\bibitem[{{Hao} {et~al.}(2010){Hao}, {McKay}, {Koester}, {Rykoff}, {Rozo},
  {Annis}, {Wechsler}, {Evrard}, {Siegel}, {Becker}, {Busha}, {Gerdes},
  {Johnston}, \& {Sheldon}}]{haokoester10}
{Hao}, J., {McKay}, T.~A., {Koester}, B.~P., {et~al.} 2010, \apjs, 191, 254

\bibitem[{{Hardcastle}(2018{\natexlab{a}})}]{hardcastlenat}
{Hardcastle}, M. 2018{\natexlab{a}}, Nature Astronomy, 2, 273

\bibitem[{{Hardcastle}(2018{\natexlab{b}})}]{Qjet_Hardcastle}
{Hardcastle}, M.~J. 2018{\natexlab{b}}, \mnras, 475, 2768

\bibitem[{{Hardcastle} \& {Croston}(2020)}]{hardcastlereview20}
{Hardcastle}, M.~J. \& {Croston}, J.~H. 2020, \nar, 88, 101539

\bibitem[{{Hardcastle} {et~al.}(2019){Hardcastle}, {Williams}, {Best},
  {Croston}, {Duncan}, {R{\"o}ttgering}, {Sabater}, {Shimwell}, {Tasse},
  {Callingham}, {Cochrane}, {de Gasperin}, {G{\"u}rkan}, {Jarvis}, {Mahatma},
  {Miley}, {Mingo}, {Mooney}, {Morabito}, {O'Sullivan}, {Prandoni},
  {Shulevski}, \& {Smith}}]{hardcastle19}
{Hardcastle}, M.~J., {Williams}, W.~L., {Best}, P.~N., {et~al.} 2019, \aap,
  622, A12

\bibitem[{{Hardee}(2007)}]{Hardee07}
{Hardee}, P.~E. 2007, \apj, 664, 26

\bibitem[{{Harwood} {et~al.}(2016){Harwood}, {Croston}, {Intema}, {Stewart},
  {Ineson}, {Hardcastle}, {Godfrey}, {Best}, {Brienza}, {Heesen}, {Mahony},
  {Morganti}, {Murgia}, {Orr{\'u}}, {R{\"o}ttgering}, {Shulevski}, \&
  {Wise}}]{Harwood16}
{Harwood}, J.~J., {Croston}, J.~H., {Intema}, H.~T., {et~al.} 2016, \mnras,
  458, 4443

\bibitem[{{Heinz} {et~al.}(2002){Heinz}, {Choi}, {Reynolds}, \&
  {Begelman}}]{heinz02}
{Heinz}, S., {Choi}, Y.-Y., {Reynolds}, C.~S., \& {Begelman}, M.~C. 2002,
  \apjl, 569, L79

\bibitem[{{Hoang} {et~al.}(2017){Hoang}, {Shimwell}, {Stroe}, {Akamatsu},
  {Brunetti}, {Donnert}, {Intema}, {Mulcahy}, {R{\"o}ttgering}, {van Weeren},
  {Bonafede}, {Br{\"u}ggen}, {Cassano}, {Chy{\.z}y}, {En{\ss}lin}, {Ferrari},
  {de Gasperin}, {Gu}, {Hoeft}, {Miley}, {Orr{\'u}}, {Pizzo}, \&
  {White}}]{duy17}
{Hoang}, D.~N., {Shimwell}, T.~W., {Stroe}, A., {et~al.} 2017, \mnras, 471,
  1107

\bibitem[{{Hocuk} \& {Barthel}(2010)}]{hocuk10}
{Hocuk}, S. \& {Barthel}, P.~D. 2010, \aap, 523, A9

\bibitem[{{Hummel} {et~al.}(1992){Hummel}, {Schalinski}, {Krichbaum}, {Rioja},
  {Quirrenbach}, {Witzel}, {Muxlow}, {Johnston}, {Matveenko}, \&
  {Shevchenko}}]{Hummel92}
{Hummel}, C.~A., {Schalinski}, C.~J., {Krichbaum}, T.~P., {et~al.} 1992, \aap,
  257, 489

\bibitem[{{Intema} {et~al.}(2017){Intema}, {Jagannathan}, {Mooley}, \&
  {Frail}}]{tgss_intema}
{Intema}, H.~T., {Jagannathan}, P., {Mooley}, K.~P., \& {Frail}, D.~A. 2017,
  \aap, 598, A78

\bibitem[{{Intema} {et~al.}(2009){Intema}, {van der Tol}, {Cotton}, {Cohen},
  {van Bemmel}, \& {R{\"o}ttgering}}]{Intema09}
{Intema}, H.~T., {van der Tol}, S., {Cotton}, W.~D., {et~al.} 2009, \aap, 501,
  1185

\bibitem[{{Ishwara-Chandra} \& {Saikia}(1999)}]{Ishwara1999}
{Ishwara-Chandra}, C.~H. \& {Saikia}, D.~J. 1999, \mnras, 309, 100

\bibitem[{{Isobe} \& {Koyama}(2015)}]{Isobe2015}
{Isobe}, N. \& {Koyama}, S. 2015, \pasj, 67, 77

\bibitem[{{Jaegers}(1987)}]{Jaegers87}
{Jaegers}, W.~J. 1987, \aaps, 71, 75

\bibitem[{{Jafelice} \& {Opher}(1992)}]{Jafelice1992}
{Jafelice}, L.~C. \& {Opher}, R. 1992, \mnras, 257, 135

\bibitem[{{Jaffe} \& {Perola}(1973)}]{JP73}
{Jaffe}, W.~J. \& {Perola}, G.~C. 1973, \aap, 26, 423

\bibitem[{{Jamrozy} {et~al.}(2008){Jamrozy}, {Konar}, {Machalski}, \&
  {Saikia}}]{Jamrozy08}
{Jamrozy}, M., {Konar}, C., {Machalski}, J., \& {Saikia}, D.~J. 2008, \mnras,
  385, 1286

\bibitem[{{Kardashev}(1962)}]{Kardashev62}
{Kardashev}, N.~S. 1962, \sovast, 6, 317

\bibitem[{{Kettenis} {et~al.}(2006){Kettenis}, {van Langevelde}, {Reynolds}, \&
  {Cotton}}]{Kettenis06}
{Kettenis}, M., {van Langevelde}, H.~J., {Reynolds}, C., \& {Cotton}, B. 2006,
  in Astronomical Society of the Pacific Conference Series, Vol. 351,
  Astronomical Data Analysis Software and Systems XV, ed. C.~{Gabriel},
  C.~{Arviset}, D.~{Ponz}, \& S.~{Enrique}, 497

\bibitem[{{Koester} {et~al.}(2007){Koester}, {McKay}, {Annis}, {Wechsler},
  {Evrard}, {Bleem}, {Becker}, {Johnston}, {Sheldon}, {Nichol}, {Miller},
  {Scranton}, {Bahcall}, {Barentine}, {Brewington}, {Brinkmann}, {Harvanek},
  {Kleinman}, {Krzesinski}, {Long}, {Nitta}, {Schneider}, {Sneddin}, {Voges},
  \& {York}}]{Koester07}
{Koester}, B.~P., {McKay}, T.~A., {Annis}, J., {et~al.} 2007, \apj, 660, 239

\bibitem[{{Komberg} \& {Pashchenko}(2009)}]{Komberg09}
{Komberg}, B.~V. \& {Pashchenko}, I.~N. 2009, Astronomy Reports, 53, 1086

\bibitem[{{Konar} {et~al.}(2008){Konar}, {Jamrozy}, {Saikia}, \&
  {}}]{Konar2008}
{Konar}, C., {Jamrozy}, M., {Saikia}, D.~J., \& {}, J. 2008, \mnras, 383, 525

\bibitem[{{Konar} {et~al.}(2004){Konar}, {Saikia}, {Ishwara-Chandra}, \&
  {Kulkarni}}]{Konar2004}
{Konar}, C., {Saikia}, D.~J., {Ishwara-Chandra}, C.~H., \& {Kulkarni}, V.~K.
  2004, \mnras, 355, 845

\bibitem[{{Kronberg}(1994)}]{kronberg94}
{Kronberg}, P.~P. 1994, Reports on Progress in Physics, 57, 325

\bibitem[{{Kronberg} {et~al.}(2004){Kronberg}, {Colgate}, {Li}, \&
  {Dufton}}]{kronberg04}
{Kronberg}, P.~P., {Colgate}, S.~A., {Li}, H., \& {Dufton}, Q.~W. 2004, \apjl,
  604, L77

\bibitem[{{Kronberg} {et~al.}(2001){Kronberg}, {Dufton}, {Li}, \&
  {Colgate}}]{Kronberg01}
{Kronberg}, P.~P., {Dufton}, Q.~W., {Li}, H., \& {Colgate}, S.~A. 2001, \apj,
  560, 178

\bibitem[{{Kroupa}(2001)}]{kroupa01}
{Kroupa}, P. 2001, \mnras, 322, 231

\bibitem[{{Ku{\'z}micz} {et~al.}(2018){Ku{\'z}micz}, {Jamrozy}, {Bronarska},
  {Janda-Boczar}, \& {Saikia}}]{Kuzmicz2018}
{Ku{\'z}micz}, A., {Jamrozy}, M., {Bronarska}, K., {Janda-Boczar}, K., \&
  {Saikia}, D.~J. 2018, \apjs, 238, 9

\bibitem[{{Ku{\'z}micz} {et~al.}(2021){Ku{\'z}micz}, {Sethi}, \&
  {Jamrozy}}]{KSJ2021}
{Ku{\'z}micz}, A., {Sethi}, S., \& {Jamrozy}, M. 2021, \apj, 922, 52

\bibitem[{{Laing} {et~al.}(1994){Laing}, {Jenkins}, {Wall}, \&
  {Unger}}]{Laing1994}
{Laing}, R.~A., {Jenkins}, C.~R., {Wall}, J.~V., \& {Unger}, S.~W. 1994, in
  Astronomical Society of the Pacific Conference Series, Vol.~54, The Physics
  of Active Galaxies, ed. G.~V. {Bicknell}, M.~A. {Dopita}, \& P.~J. {Quinn},
  201

\bibitem[{{Lara} {et~al.}(2001){Lara}, {Cotton}, {Feretti}, {Giovannini},
  {Marcaide}, {M{\'a}rquez}, \& {Venturi}}]{lara2001}
{Lara}, L., {Cotton}, W.~D., {Feretti}, L., {et~al.} 2001, \aap, 370, 409

\bibitem[{{Leahy}(1991)}]{Leahy1991book}
{Leahy}, J.~P. 1991, in Beams and Jets in Astrophysics, Vol.~19, 100

\bibitem[{{Lin} {et~al.}(2018){Lin}, {Huang}, \& {Chen}}]{Lin18}
{Lin}, Y.-T., {Huang}, H.-J., \& {Chen}, Y.-C. 2018, \aj, 155, 188

\bibitem[{{Lobanov} \& {Zensus}(2001)}]{Lobanov01}
{Lobanov}, A.~P. \& {Zensus}, J.~A. 2001, Science, 294, 128

\bibitem[{{Longair} {et~al.}(1973){Longair}, {Ryle}, \& {Scheuer}}]{longair73}
{Longair}, M.~S., {Ryle}, M., \& {Scheuer}, P.~A.~G. 1973, \mnras, 164, 243

\bibitem[{{Machalski} {et~al.}(2009){Machalski}, {Jamrozy}, \&
  {Saikia}}]{machalski09}
{Machalski}, J., {Jamrozy}, M., \& {Saikia}, D.~J. 2009, \mnras, 395, 812

\bibitem[{{Mack} {et~al.}(1997){Mack}, {Klein}, {O'Dea}, \& {Willis}}]{mack97}
{Mack}, K.~H., {Klein}, U., {O'Dea}, C.~P., \& {Willis}, A.~G. 1997, \aaps,
  123, 423

\bibitem[{{Mack} {et~al.}(1998){Mack}, {Klein}, {O'Dea}, {Willis}, \&
  {Saripalli}}]{mack98}
{Mack}, K.~H., {Klein}, U., {O'Dea}, C.~P., {Willis}, A.~G., \& {Saripalli}, L.
  1998, \aap, 329, 431

\bibitem[{{Mahato} {et~al.}(2022){Mahato}, {Dabhade}, {Saikia}, {Combes},
  {Bagchi}, {Ho}, \& {Raychaudhury}}]{sagan3}
{Mahato}, M., {Dabhade}, P., {Saikia}, D.~J., {et~al.} 2022, \aap, 660, A59

\bibitem[{{Malarecki} {et~al.}(2015){Malarecki}, {Jones}, {Saripalli},
  {Staveley-Smith}, \& {Subrahmanyan}}]{malarecki15}
{Malarecki}, J.~M., {Jones}, D.~H., {Saripalli}, L., {Staveley-Smith}, L., \&
  {Subrahmanyan}, R. 2015, \mnras, 449, 955

\bibitem[{{Marshall} {et~al.}(2001){Marshall}, {Harris}, {Grimes}, {Drake},
  {Fruscione}, {Juda}, {Kraft}, {Mathur}, {Murray}, {Ogle}, {Pease},
  {Schwartz}, {Siemiginowska}, {Vrtilek}, \& {Wargelin}}]{Marshall01}
{Marshall}, H.~L., {Harris}, D.~E., {Grimes}, J.~P., {et~al.} 2001, \apjl, 549,
  L167

\bibitem[{{McNamara} \& {Nulsen}(2007)}]{McNamara07}
{McNamara}, B.~R. \& {Nulsen}, P.~E.~J. 2007, \araa, 45, 117

\bibitem[{{McNamara} {et~al.}(2005){McNamara}, {Nulsen}, {Wise}, {Rafferty},
  {Carilli}, {Sarazin}, \& {Blanton}}]{McNamara05}
{McNamara}, B.~R., {Nulsen}, P.~E.~J., {Wise}, M.~W., {et~al.} 2005, \nat, 433,
  45

\bibitem[{{Mertens} {et~al.}(2016){Mertens}, {Lobanov}, {Walker}, \&
  {Hardee}}]{Mertens16}
{Mertens}, F., {Lobanov}, A.~P., {Walker}, R.~C., \& {Hardee}, P.~E. 2016,
  \aap, 595, A54

\bibitem[{{Meyer} {et~al.}(2020){Meyer}, {Petropoulou}, \&
  {Christie}}]{Meyer2020}
{Meyer}, M., {Petropoulou}, M., \& {Christie}, I. 2020, arXiv e-prints,
  arXiv:2012.09944

\bibitem[{{Miley}(1980)}]{Miley}
{Miley}, G. 1980, \araa, 18, 165

\bibitem[{{Mingo} {et~al.}(2022){Mingo}, {Croston}, {Best}, {Duncan},
  {Hardcastle}, {Kondapally}, {Prandoni}, {Sabater}, {Shimwell}, {Williams},
  {Baldi}, {Bonato}, {Bondi}, {Dabhade}, {G{\"u}rkan}, {Ineson},
  {Magliocchetti}, {Miley}, {Pierce}, \& {R{\"o}ttgering}}]{Mingo22}
{Mingo}, B., {Croston}, J.~H., {Best}, P.~N., {et~al.} 2022, \mnras, 511, 3250

\bibitem[{{Mingo} {et~al.}(2019){Mingo}, {Croston}, {Hardcastle}, {Best},
  {Duncan}, {Morganti}, {Rottgering}, {Sabater}, {Shimwell}, {Williams},
  {Brienza}, {Gurkan}, {Mahatma}, {Morabito}, {Prandoni}, {Bondi}, {Ineson}, \&
  {Mooney}}]{mingo19}
{Mingo}, B., {Croston}, J.~H., {Hardcastle}, M.~J., {et~al.} 2019, \mnras, 488,
  2701

\bibitem[{{Mingo} {et~al.}(2014){Mingo}, {Hardcastle}, {Croston}, {Dicken},
  {Evans}, {Morganti}, \& {Tadhunter}}]{mingo14}
{Mingo}, B., {Hardcastle}, M.~J., {Croston}, J.~H., {et~al.} 2014, \mnras, 440,
  269

\bibitem[{{Mizuno} {et~al.}(2014){Mizuno}, {Hardee}, \& {Nishikawa}}]{Mizuno14}
{Mizuno}, Y., {Hardee}, P.~E., \& {Nishikawa}, K.-I. 2014, \apj, 784, 167

\bibitem[{{Mizuno} {et~al.}(2009){Mizuno}, {Lyubarsky}, {Nishikawa}, \&
  {Hardee}}]{Mizuno09}
{Mizuno}, Y., {Lyubarsky}, Y., {Nishikawa}, K.-I., \& {Hardee}, P.~E. 2009,
  \apj, 700, 684

\bibitem[{{Mizuno} {et~al.}(2012){Mizuno}, {Lyubarsky}, {Nishikawa}, \&
  {Hardee}}]{Mizuno12}
{Mizuno}, Y., {Lyubarsky}, Y., {Nishikawa}, K.-I., \& {Hardee}, P.~E. 2012,
  \apj, 757, 16

\bibitem[{{Morton}(1991)}]{Vac2air-M91}
{Morton}, D.~C. 1991, \apjs, 77, 119

\bibitem[{{Mukherjee} {et~al.}(2020){Mukherjee}, {Bodo}, {Mignone}, {Rossi}, \&
  {Vaidya}}]{Dipanjan21p1}
{Mukherjee}, D., {Bodo}, G., {Mignone}, A., {Rossi}, P., \& {Vaidya}, B. 2020,
  \mnras, 499, 681

\bibitem[{{Mukherjee} {et~al.}(2021){Mukherjee}, {Bodo}, {Rossi}, {Mignone}, \&
  {Vaidya}}]{Dipanjan21p2}
{Mukherjee}, D., {Bodo}, G., {Rossi}, P., {Mignone}, A., \& {Vaidya}, B. 2021,
  \mnras, 505, 2267

\bibitem[{{Nakamura} {et~al.}(2007){Nakamura}, {Li}, \& {Li}}]{nakamura07}
{Nakamura}, M., {Li}, H., \& {Li}, S. 2007, \apj, 656, 721

\bibitem[{{Nakamura} \& {Meier}(2004)}]{nakamura04}
{Nakamura}, M. \& {Meier}, D.~L. 2004, \apj, 617, 123

\bibitem[{{Nakamura} {et~al.}(2001){Nakamura}, {Uchida}, \&
  {Hirose}}]{Nakamura01}
{Nakamura}, M., {Uchida}, Y., \& {Hirose}, S. 2001, \na, 6, 61

\bibitem[{{Noordam}(2004)}]{Noordam04}
{Noordam}, J.~E. 2004, in Society of Photo-Optical Instrumentation Engineers
  (SPIE) Conference Series, Vol. 5489, Ground-based Telescopes, ed.
  J.~{Oschmann}, Jacobus~M., 817--825

\bibitem[{{Offringa} {et~al.}(2014){Offringa}, {McKinley}, {Hurley-Walker},
  {Briggs}, {Wayth}, {Kaplan}, {Bell}, {Feng}, {Neben}, {Hughes}, {Rhee},
  {Murphy}, {Bhat}, {Bernardi}, {Bowman}, {Cappallo}, {Corey}, {Deshpand e},
  {Emrich}, {Ewall-Wice}, {Gaensler}, {Goeke}, {Greenhill}, {Hazelton},
  {Hindson}, {Johnston-Hollitt}, {Jacobs}, {Kasper}, {Kratzenberg}, {Lenc},
  {Lonsdale}, {Lynch}, {McWhirter}, {Mitchell}, {Morales}, {Morgan},
  {Kudryavtseva}, {Oberoi}, {Ord}, {Pindor}, {Procopio}, {Prabu}, {Riding},
  {Roshi}, {Shankar}, {Srivani}, {Subrahmanyan}, {Tingay}, {Waterson},
  {Webster}, {Whitney}, {Williams}, \& {Williams}}]{Offringa14wsclean}
{Offringa}, A.~R., {McKinley}, B., {Hurley-Walker}, N., {et~al.} 2014, \mnras,
  444, 606

\bibitem[{{Pacholczyk}(1970)}]{Pacholczyk}
{Pacholczyk}, A.~G. 1970, {Radio astrophysics. Nonthermal processes in galactic
  and extragalactic sources}

\bibitem[{Pasetto {et~al.}(2021)Pasetto, Carrasco-Gonz{\'{a}}lez, G{\'{o}}mez,
  Mart{\'{\i}}, Perucho, O'Sullivan, Anderson, D{\'{\i}}az-Gonz{\'{a}}lez,
  Fuentes, \& Wardle}]{Pasetto_2021}
Pasetto, A., Carrasco-Gonz{\'{a}}lez, C., G{\'{o}}mez, J.~L., {et~al.} 2021,
  The Astrophysical Journal Letters, 923, L5

\bibitem[{{Paturel} {et~al.}(2003){Paturel}, {Petit}, {Prugniel}, {Theureau},
  {Rousseau}, {Brouty}, {Dubois}, \& {Cambr{\'e}sy}}]{HYPERLEDA}
{Paturel}, G., {Petit}, C., {Prugniel}, P., {et~al.} 2003, \aap, 412, 45

\bibitem[{{Peacock}(1981)}]{Peacock81}
{Peacock}, J.~A. 1981, \mnras, 196, 135

\bibitem[{{Perley} {et~al.}(1984){Perley}, {Bridle}, \& {Willis}}]{Perley84}
{Perley}, R.~A., {Bridle}, A.~H., \& {Willis}, A.~G. 1984, \apjs, 54, 291

\bibitem[{{Perucho} {et~al.}(2007){Perucho}, {Hanasz}, {Mart{\'\i}}, \&
  {Miralles}}]{Perucho07}
{Perucho}, M., {Hanasz}, M., {Mart{\'\i}}, J.-M., \& {Miralles}, J.-A. 2007,
  \pre, 75, 056312

\bibitem[{{Pirya} {et~al.}(2012){Pirya}, {Saikia}, {Singh}, \&
  {Chandola}}]{Pirya}
{Pirya}, A., {Saikia}, D.~J., {Singh}, M., \& {Chandola}, H.~C. 2012, \mnras,
  426, 758

\bibitem[{{Planck Collaboration} {et~al.}(2016){Planck Collaboration}, {Ade},
  {Aghanim}, {Arnaud}, {Ashdown}, {Aumont}, {Baccigalupi}, {Banday},
  {Barreiro}, {Bartlett}, {Bartolo}, {Battaner}, {Battye}, {Benabed},
  {Beno{\^\i}t}, {Benoit-L{\'e}vy}, {Bernard}, {Bersanelli}, {Bielewicz},
  {Bock}, {Bonaldi}, {Bonavera}, {Bond}, {Borrill}, {Bouchet}, {Boulanger},
  {Bucher}, {Burigana}, {Butler}, {Calabrese}, {Cardoso}, {Catalano},
  {Challinor}, {Chamballu}, {Chary}, {Chiang}, {Chluba}, {Christensen},
  {Church}, {Clements}, {Colombi}, {Colombo}, {Combet}, {Coulais}, {Crill},
  {Curto}, {Cuttaia}, {Danese}, {Davies}, {Davis}, {de Bernardis}, {de Rosa},
  {de Zotti}, {Delabrouille}, {D{\'e}sert}, {Di Valentino}, {Dickinson},
  {Diego}, {Dolag}, {Dole}, {Donzelli}, {Dor{\'e}}, {Douspis}, {Ducout},
  {Dunkley}, {Dupac}, {Efstathiou}, {Elsner}, {En{\ss}lin}, {Eriksen},
  {Farhang}, {Fergusson}, {Finelli}, {Forni}, {Frailis}, {Fraisse},
  {Franceschi}, {Frejsel}, {Galeotta}, {Galli}, {Ganga}, {Gauthier}, {Gerbino},
  {Ghosh}, {Giard}, {Giraud-H{\'e}raud}, {Giusarma}, {Gjerl{\o}w},
  {Gonz{\'a}lez-Nuevo}, {G{\'o}rski}, {Gratton}, {Gregorio}, {Gruppuso},
  {Gudmundsson}, {Hamann}, {Hansen}, {Hanson}, {Harrison}, {Helou}, {Henrot-
  Versill{\'e}}, {Hern{\'a}ndez-Monteagudo}, {Herranz}, {Hildebrandt}, {Hivon},
  {Hobson}, {Holmes}, {Hornstrup}, {Hovest}, {Huang}, {Huffenberger}, {Hurier},
  {Jaffe}, {Jaffe}, {Jones}, {Juvela}, {Keih{\"a}nen}, {Keskitalo}, {Kisner},
  {Kneissl}, {Knoche}, {Knox}, {Kunz}, {Kurki-Suonio}, {Lagache},
  {L{\"a}hteenm{\"a}ki}, {Lamarre}, {Lasenby}, {Lattanzi}, {Lawrence}, {Leahy},
  {Leonardi}, {Lesgourgues}, {Levrier}, {Lewis}, {Liguori}, {Lilje},
  {Linden-V{\o}rnle}, {L{\'o}pez-Caniego}, {Lubin}, {Mac{\'\i}as-P{\'e}rez},
  {Maggio}, {Maino}, {Mandolesi}, {Mangilli}, {Marchini}, {Maris}, {Martin},
  {Martinelli}, {Mart{\'\i}nez-Gonz{\'a}lez}, {Masi}, {Matarrese}, {McGehee},
  {Meinhold}, {Melchiorri}, {Melin}, {Mendes}, {Mennella}, {Migliaccio},
  {Millea}, {Mitra}, {Miville-Desch{\^e}nes}, {Moneti}, {Montier}, {Morgante},
  {Mortlock}, {Moss}, {Munshi}, {Murphy}, {Naselsky}, {Nati}, {Natoli},
  {Netterfield}, {N{\o}rgaard-Nielsen}, {Noviello}, {Novikov}, {Novikov},
  {Oxborrow}, {Paci}, {Pagano}, {Pajot}, {Paladini}, {Paoletti}, {Partridge},
  {Pasian}, {Patanchon}, {Pearson}, {Perdereau}, {Perotto}, {Perrotta},
  {Pettorino}, {Piacentini}, {Piat}, {Pierpaoli}, {Pietrobon}, {Plaszczynski},
  {Pointecouteau}, {Polenta}, {Popa}, {Pratt}, {Pr{\'e}zeau}, {Prunet},
  {Puget}, {Rachen}, {Reach}, {Rebolo}, {Reinecke}, {Remazeilles}, {Renault},
  {Renzi}, {Ristorcelli}, {Rocha}, {Rosset}, {Rossetti}, {Roudier},
  {Rouill{\'e} d'Orfeuil}, {Rowan-Robinson}, {Rubi{\~n}o-Mart{\'\i}n},
  {Rusholme}, {Said}, {Salvatelli}, {Salvati}, {Sandri}, {Santos},
  {Savelainen}, {Savini}, {Scott}, {Seiffert}, {Serra}, {Shellard}, {Spencer},
  {Spinelli}, {Stolyarov}, {Stompor}, {Sudiwala}, {Sunyaev}, {Sutton},
  {Suur-Uski}, {Sygnet}, {Tauber}, {Terenzi}, {Toffolatti}, {Tomasi},
  {Tristram}, {Trombetti}, {Tucci}, {Tuovinen}, {T{\"u}rler}, {Umana},
  {Valenziano}, {Valiviita}, {Van Tent}, {Vielva}, {Villa}, {Wade}, {Wandelt},
  {Wehus}, {White}, {White}, {Wilkinson}, {Yvon}, {Zacchei}, \&
  {Zonca}}]{2016A&A...594A..13P}
{Planck Collaboration}, {Ade}, P.~A.~R., {Aghanim}, N., {et~al.} 2016, \aap,
  594, A13

\bibitem[{{Quilis} {et~al.}(2001){Quilis}, {Bower}, \& {Balogh}}]{quilis01}
{Quilis}, V., {Bower}, R.~G., \& {Balogh}, M.~L. 2001, \mnras, 328, 1091

\bibitem[{{Rees}(1987)}]{Rees1987}
{Rees}, M.~J. 1987, \qjras, 28, 197

\bibitem[{{Riley}(1975)}]{Riley75}
{Riley}, J.~M. 1975, \mnras, 170, 53

\bibitem[{{Robitaille} \& {Bressert}(2012)}]{apl}
{Robitaille}, T. \& {Bressert}, E. 2012, {APLpy: Astronomical Plotting Library
  in Python}

\bibitem[{{Romanova} \& {Lovelace}(1992)}]{Romanova92}
{Romanova}, M.~M. \& {Lovelace}, R.~V.~E. 1992, \aap, 262, 26

\bibitem[{{Ruzmaikin} {et~al.}(1989){Ruzmaikin}, {Sokolov}, \&
  {Shukurov}}]{Ruzmaikin1989}
{Ruzmaikin}, A., {Sokolov}, D., \& {Shukurov}, A. 1989, \mnras, 241, 1

\bibitem[{{Sadat} {et~al.}(2004){Sadat}, {Blanchard}, {Kneib}, {Mathez},
  {Madore}, \& {Mazzarella}}]{Sadat2004}
{Sadat}, R., {Blanchard}, A., {Kneib}, J.~P., {et~al.} 2004, \aap, 424, 1097

\bibitem[{{Safouris} {et~al.}(2009){Safouris}, {Subrahmanyan}, {Bicknell}, \&
  {Saripalli}}]{Safouris09}
{Safouris}, V., {Subrahmanyan}, R., {Bicknell}, G.~V., \& {Saripalli}, L. 2009,
  \mnras, 393, 2

\bibitem[{{Saikia}(2022)}]{Saikia22}
{Saikia}, D.~J. 2022, Journal of Astrophysics and Astronomy, arXiv:2206.05803

\bibitem[{{Saripalli} {et~al.}(2005){Saripalli}, {Hunstead}, {Subrahmanyan}, \&
  {Boyce}}]{Saripalli2005}
{Saripalli}, L., {Hunstead}, R.~W., {Subrahmanyan}, R., \& {Boyce}, E. 2005,
  \aj, 130, 896

\bibitem[{{Saxena} {et~al.}(2019){Saxena}, {R{\"o}ttgering}, {Duncan}, {Hill},
  {Best}, {Indahl}, {Marinello}, {Overzier}, {Pentericci}, {Prandoni},
  {Dannerbauer}, \& {Barrena}}]{Saxena19}
{Saxena}, A., {R{\"o}ttgering}, H.~J.~A., {Duncan}, K.~J., {et~al.} 2019,
  \mnras, 489, 5053

\bibitem[{{Scaife} \& {Heald}(2012)}]{ScaifeHeald12}
{Scaife}, A. M.~M. \& {Heald}, G.~H. 2012, \mnras, 423, L30

\bibitem[{{Scheuer}(1974)}]{Scheuer74}
{Scheuer}, P.~A.~G. 1974, \mnras, 166, 513

\bibitem[{{Scheuer} \& {Williams}(1968)}]{scheuer68}
{Scheuer}, P.~A.~G. \& {Williams}, P.~J.~S. 1968, \araa, 6, 321

\bibitem[{{Schoenmakers} {et~al.}(2000){Schoenmakers}, {Mack}, {de Bruyn},
  {R{\"o}ttgering}, {Klein}, \& {van der Laan}}]{Schoenmakers_spec}
{Schoenmakers}, A.~P., {Mack}, K.~H., {de Bruyn}, A.~G., {et~al.} 2000, \aaps,
  146, 293

\bibitem[{{Shimwell} {et~al.}(2019){Shimwell}, {Tasse}, {Hardcastle}, {Mechev},
  {Williams}, {Best}, {R{\"o}ttgering}, {Callingham}, {Dijkema}, {de Gasperin},
  {Hoang}, {Hugo}, {Mirmont}, {Oonk}, {Prandoni}, {Rafferty}, {Sabater},
  {Smirnov}, {van Weeren}, {White}, {Atemkeng}, {Bester}, {Bonnassieux},
  {Br{\"u}ggen}, {Brunetti}, {Chy{\.z}y}, {Cochrane}, {Conway}, {Croston},
  {Danezi}, {Duncan}, {Haverkorn}, {Heald}, {Iacobelli}, {Intema}, {Jackson},
  {Jamrozy}, {Jarvis}, {Lakhoo}, {Mevius}, {Miley}, {Morabito}, {Morganti},
  {Nisbet}, {Orr{\'u}}, {Perkins}, {Pizzo}, {Schrijvers}, {Smith}, {Vermeulen},
  {Wise}, {Alegre}, {Bacon}, {van Bemmel}, {Beswick}, {Bonafede}, {Botteon},
  {Bourke}, {Brienza}, {Calistro Rivera}, {Cassano}, {Clarke}, {Conselice},
  {Dettmar}, {Drabent}, {Dumba}, {Emig}, {En{\ss}lin}, {Ferrari}, {Garrett},
  {G{\'e}nova-Santos}, {Goyal}, {G{\"u}rkan}, {Hale}, {Harwood}, {Heesen},
  {Hoeft}, {Horellou}, {Jackson}, {Kokotanekov}, {Kondapally},
  {Kunert-Bajraszewska}, {Mahatma}, {Mahony}, {Mandal}, {McKean}, {Merloni},
  {Mingo}, {Miskolczi}, {Mooney}, {Nikiel-Wroczy{\'n}ski}, {O'Sullivan},
  {Quinn}, {Reich}, {Roskowi{\'n}ski}, {Rowlinson}, {Savini}, {Saxena},
  {Schwarz}, {Shulevski}, {Sridhar}, {Stacey}, {Urquhart}, {van der Wiel},
  {Varenius}, {Webster}, \& {Wilber}}]{lotssshimwell}
{Shimwell}, T.~W., {Tasse}, C., {Hardcastle}, M.~J., {et~al.} 2019, \aap, 622,
  A1

\bibitem[{{Shukla} \& {Mannheim}(2020)}]{shukla2020}
{Shukla}, A. \& {Mannheim}, K. 2020, Nature Communications, 11, 4176

\bibitem[{{Shulevski} {et~al.}(2019){Shulevski}, {Barthel}, {Morganti},
  {Harwood}, {Brienza}, {Shimwell}, {R{\"o}ttgering}, {White}, {Callingham},
  {Mooney}, \& {Rafferty}}]{Shulevski19}
{Shulevski}, A., {Barthel}, P.~D., {Morganti}, R., {et~al.} 2019, \aap, 628,
  A69

\bibitem[{{Simonte} {et~al.}(2022){Simonte}, {Andernach}, {Br{\"u}ggen},
  {Schwarz}, {Prandoni}, \& {Willis}}]{Simonte2022}
{Simonte}, M., {Andernach}, H., {Br{\"u}ggen}, M., {et~al.} 2022, \mnras, 515,
  2032

\bibitem[{{Sironi} {et~al.}(2015){Sironi}, {Petropoulou}, \&
  {Giannios}}]{sironi15}
{Sironi}, L., {Petropoulou}, M., \& {Giannios}, D. 2015, \mnras, 450, 183

\bibitem[{{Sironi} {et~al.}(2013){Sironi}, {Spitkovsky}, \& {Arons}}]{Sironi13}
{Sironi}, L., {Spitkovsky}, A., \& {Arons}, J. 2013, \apj, 771, 54

\bibitem[{{Smirnov} \& {Tasse}(2015)}]{Smirnov15}
{Smirnov}, O.~M. \& {Tasse}, C. 2015, \mnras, 449, 2668

\bibitem[{{Subrahmanyan} {et~al.}(1996){Subrahmanyan}, {Saripalli}, \&
  {Hunstead}}]{ravi96}
{Subrahmanyan}, R., {Saripalli}, L., \& {Hunstead}, R.~W. 1996, \mnras, 279,
  257

\bibitem[{{Subrahmanyan} {et~al.}(2008){Subrahmanyan}, {Saripalli}, {Safouris},
  \& {Hunstead}}]{ravi08}
{Subrahmanyan}, R., {Saripalli}, L., {Safouris}, V., \& {Hunstead}, R.~W. 2008,
  \apj, 677, 63

\bibitem[{{Swarup}(1991)}]{Swarup1991GMRT1}
{Swarup}, G. 1991, in Astronomical Society of the Pacific Conference Series,
  Vol.~19, IAU Colloq. 131: Radio Interferometry. Theory, Techniques, and
  Applications, ed. T.~J. {Cornwell} \& R.~A. {Perley}, 376--380

\bibitem[{{Swarup} {et~al.}(1991){Swarup}, {Ananthakrishnan}, {Kapahi}, {Rao},
  {Subrahmanya}, \& {Kulkarni}}]{Swarup1991GMRT2}
{Swarup}, G., {Ananthakrishnan}, S., {Kapahi}, V.~K., {et~al.} 1991, Current
  Science, 60, 95

\bibitem[{{Tasse}(2014)}]{Tasse14}
{Tasse}, C. 2014, \aap, 566, A127

\bibitem[{{Tasse} {et~al.}(2018){Tasse}, {Hugo}, {Mirmont}, {Smirnov},
  {Atemkeng}, {Bester}, {Hardcastle}, {Lakhoo}, {Perkins}, \&
  {Shimwell}}]{Tasse18}
{Tasse}, C., {Hugo}, B., {Mirmont}, M., {et~al.} 2018, \aap, 611, A87

\bibitem[{{Tasse} {et~al.}(2021){Tasse}, {Shimwell}, {Hardcastle},
  {O'Sullivan}, {van Weeren}, {Best}, {Bester}, {Hugo}, {Smirnov}, {Sabater},
  {Calistro-Rivera}, {de Gasperin}, {Morabito}, {R{\"o}ttgering}, {Williams},
  {Bonato}, {Bondi}, {Botteon}, {Br{\"u}ggen}, {Brunetti}, {Chy{\.z}y},
  {Garrett}, {G{\"u}rkan}, {Jarvis}, {Kondapally}, {Mandal}, {Prandoni},
  {Repetti}, {Retana-Montenegro}, {Schwarz}, {Shulevski}, \& {Wiaux}}]{Tasse20}
{Tasse}, C., {Shimwell}, T., {Hardcastle}, M.~J., {et~al.} 2021, \aap, 648, A1

\bibitem[{{Tomimatsu}(1994)}]{Tomimatsu1994}
{Tomimatsu}, A. 1994, \pasj, 46, 123

\bibitem[{{van der Laan} \& {Perola}(1969)}]{vanderLaan1969}
{van der Laan}, H. \& {Perola}, G.~C. 1969, \aap, 3, 468

\bibitem[{{van Diepen} {et~al.}(2018){van Diepen}, {Dijkema}, \&
  {Offringa}}]{dppp}
{van Diepen}, G., {Dijkema}, T.~J., \& {Offringa}, A. 2018, ascl:1804.003

\bibitem[{{van Haarlem} {et~al.}(2013){van Haarlem}, {Wise}, {Gunst}, {Heald},
  {McKean}, {Hessels}, {de Bruyn}, {Nijboer}, {Swinbank}, {Fallows},
  {Brentjens}, {Nelles}, {Beck}, {Falcke}, {Fender}, {H{\"o}randel},
  {Koopmans}, {Mann}, {Miley}, {R{\"o}ttgering}, {Stappers}, {Wijers},
  {Zaroubi}, {van den Akker}, {Alexov}, {Anderson}, {Anderson}, {van Ardenne},
  {Arts}, {Asgekar}, {Avruch}, {Batejat}, {B{\"a}hren}, {Bell}, {Bell}, {van
  Bemmel}, {Bennema}, {Bentum}, {Bernardi}, {Best}, {B{\^i}rzan}, {Bonafede},
  {Boonstra}, {Braun}, {Bregman}, {Breitling}, {van de Brink}, {Broderick},
  {Broekema}, {Brouw}, {Br{\"u}ggen}, {Butcher}, {van Cappellen}, {Ciardi},
  {Coenen}, {Conway}, {Coolen}, {Corstanje}, {Damstra}, {Davies}, {Deller},
  {Dettmar}, {van Diepen}, {Dijkstra}, {Donker}, {Doorduin}, {Dromer}, {Drost},
  {van Duin}, {Eisl{\"o}ffel}, {van Enst}, {Ferrari}, {Frieswijk}, {Gankema},
  {Garrett}, {de Gasperin}, {Gerbers}, {de Geus}, {Grie{\ss}meier}, {Grit},
  {Gruppen}, {Hamaker}, {Hassall}, {Hoeft}, {Holties}, {Horneffer}, {van der
  Horst}, {van Houwelingen}, {Huijgen}, {Iacobelli}, {Intema}, {Jackson},
  {Jelic}, {de Jong}, {Juette}, {Kant}, {Karastergiou}, {Koers}, {Kollen},
  {Kondratiev}, {Kooistra}, {Koopman}, {Koster}, {Kuniyoshi}, {Kramer},
  {Kuper}, {Lambropoulos}, {Law}, {van Leeuwen}, {Lemaitre}, {Loose}, {Maat},
  {Macario}, {Markoff}, {Masters}, {McFadden}, {McKay-Bukowski}, {Meijering},
  {Meulman}, {Mevius}, {Middelberg}, {Millenaar}, {Miller-Jones}, {Mohan},
  {Mol}, {Morawietz}, {Morganti}, {Mulcahy}, {Mulder}, {Munk}, {Nieuwenhuis},
  {van Nieuwpoort}, {Noordam}, {Norden}, {Noutsos}, {Offringa}, {Olofsson},
  {Omar}, {Orr{\'u}}, {Overeem}, {Paas}, {Pandey-Pommier}, {Pandey}, {Pizzo},
  {Polatidis}, {Rafferty}, {Rawlings}, {Reich}, {de Reijer}, {Reitsma},
  {Renting}, {Riemers}, {Rol}, {Romein}, {Roosjen}, {Ruiter}, {Scaife}, {van
  der Schaaf}, {Scheers}, {Schellart}, {Schoenmakers}, {Schoonderbeek},
  {Serylak}, {Shulevski}, {Sluman}, {Smirnov}, {Sobey}, {Spreeuw}, {Steinmetz},
  {Sterks}, {Stiepel}, {Stuurwold}, {Tagger}, {Tang}, {Tasse}, {Thomas},
  {Thoudam}, {Toribio}, {van der Tol}, {Usov}, {van Veelen}, {van der Veen},
  {ter Veen}, {Verbiest}, {Vermeulen}, {Vermaas}, {Vocks}, {Vogt}, {de Vos},
  {van der Wal}, {van Weeren}, {Weggemans}, {Weltevrede}, {White}, {Wijnholds},
  {Wilhelmsson}, {Wucknitz}, {Yatawatta}, {Zarka}, {Zensus}, \& {van
  Zwieten}}]{lofar}
{van Haarlem}, M.~P., {Wise}, M.~W., {Gunst}, A.~W., {et~al.} 2013, \aap, 556,
  A2

\bibitem[{{van Weeren} {et~al.}(2021){van Weeren}, {Shimwell}, {Botteon},
  {Brunetti}, {Br{\"u}ggen}, {Boxelaar}, {Cassano}, {Di Gennaro},
  {Andrade-Santos}, {Bonnassieux}, {Bonafede}, {Cuciti}, {Dallacasa}, {de
  Gasperin}, {Gastaldello}, {Hardcastle}, {Hoeft}, {Kraft}, {Mandal},
  {Rossetti}, {R{\"o}ttgering}, {Tasse}, \& {Wilber}}]{vanWeeren20}
{van Weeren}, R.~J., {Shimwell}, T.~W., {Botteon}, A., {et~al.} 2021, \aap,
  651, A115

\bibitem[{{van Weeren} {et~al.}(2016){van Weeren}, {Williams}, {Hardcastle},
  {Shimwell}, {Rafferty}, {Sabater}, {Heald}, {Sridhar}, {Dijkema}, {Brunetti},
  {Br{\"u}ggen}, {Andrade-Santos}, {Ogrean}, {R{\"o}ttgering}, {Dawson},
  {Forman}, {de Gasperin}, {Jones}, {Miley}, {Rudnick}, {Sarazin}, {Bonafede},
  {Best}, {B{\^\i}rzan}, {Cassano}, {Chy{\.z}y}, {Croston}, {Ensslin},
  {Ferrari}, {Hoeft}, {Horellou}, {Jarvis}, {Kraft}, {Mevius}, {Intema},
  {Murray}, {Orr{\'u}}, {Pizzo}, {Simionescu}, {Stroe}, {van der Tol}, \&
  {White}}]{vanWeeren2016}
{van Weeren}, R.~J., {Williams}, W.~L., {Hardcastle}, M.~J., {et~al.} 2016,
  \apjs, 223, 2

\bibitem[{{Waggett} {et~al.}(1977){Waggett}, {Warner}, \&
  {Baldwin}}]{Waggett77}
{Waggett}, P.~C., {Warner}, P.~J., \& {Baldwin}, J.~E. 1977, \mnras, 181, 465

\bibitem[{{Wen} {et~al.}(2012){Wen}, {Han}, \& {Liu}}]{whl12}
{Wen}, Z.~L., {Han}, J.~L., \& {Liu}, F.~S. 2012, The Astrophysical Journal
  Supplement Series, 199, 34

\bibitem[{{Williams} {et~al.}(2016){Williams}, {van Weeren}, {R{\"o}ttgering},
  {Best}, {Dijkema}, {de Gasperin}, {Hardcastle}, {Heald}, {Prand oni},
  {Sabater}, {Shimwell}, {Tasse}, {van Bemmel}, {Br{\"u}ggen}, {Brunetti},
  {Conway}, {En{\ss}lin}, {Engels}, {Falcke}, {Ferrari}, {Haverkorn},
  {Jackson}, {Jarvis}, {Kapi{\'n}ska}, {Mahony}, {Miley}, {Morabito},
  {Morganti}, {Orr{\'u}}, {Retana-Montenegro}, {Sridhar}, {Toribio}, {White},
  {Wise}, \& {Zwart}}]{Williams2016}
{Williams}, W.~L., {van Weeren}, R.~J., {R{\"o}ttgering}, H.~J.~A., {et~al.}
  2016, \mnras, 460, 2385

\bibitem[{{Worrall} {et~al.}(2007){Worrall}, {Birkinshaw}, {Laing}, {Cotton},
  \& {Bridle}}]{worrall07}
{Worrall}, D.~M., {Birkinshaw}, M., {Laing}, R.~A., {Cotton}, W.~D., \&
  {Bridle}, A.~H. 2007, \mnras, 380, 2

\bibitem[{{York} {et~al.}(2000){York}, {Adelman}, {Anderson}, {Anderson},
  {Annis}, {Bahcall}, {Bakken}, {Barkhouser}, {Bastian}, {Berman}, {Boroski},
  {Bracker}, {Briegel}, {Briggs}, {Brinkmann}, {Brunner}, {Burles}, {Carey},
  {Carr}, {Castander}, {Chen}, {Colestock}, {Connolly}, {Crocker}, {Csabai},
  {Czarapata}, {Davis}, {Doi}, {Dombeck}, {Eisenstein}, {Ellman}, {Elms},
  {Evans}, {Fan}, {Federwitz}, {Fiscelli}, {Friedman}, {Frieman}, {Fukugita},
  {Gillespie}, {Gunn}, {Gurbani}, {de Haas}, {Haldeman}, {Harris}, {Hayes},
  {Heckman}, {Hennessy}, {Hindsley}, {Holm}, {Holmgren}, {Huang}, {Hull},
  {Husby}, {Ichikawa}, {Ichikawa}, {Ivezi{\'c}}, {Kent}, {Kim}, {Kinney},
  {Klaene}, {Kleinman}, {Kleinman}, {Knapp}, {Korienek}, {Kron}, {Kunszt},
  {Lamb}, {Lee}, {Leger}, {Limmongkol}, {Lindenmeyer}, {Long}, {Loomis},
  {Loveday}, {Lucinio}, {Lupton}, {MacKinnon}, {Mannery}, {Mantsch}, {Margon},
  {McGehee}, {McKay}, {Meiksin}, {Merelli}, {Monet}, {Munn}, {Narayanan},
  {Nash}, {Neilsen}, {Neswold}, {Newberg}, {Nichol}, {Nicinski}, {Nonino},
  {Okada}, {Okamura}, {Ostriker}, {Owen}, {Pauls}, {Peoples}, {Peterson},
  {Petravick}, {Pier}, {Pope}, {Pordes}, {Prosapio}, {Rechenmacher}, {Quinn},
  {Richards}, {Richmond}, {Rivetta}, {Rockosi}, {Ruthmansdorfer}, {Sandford},
  {Schlegel}, {Schneider}, {Sekiguchi}, {Sergey}, {Shimasaku}, {Siegmund},
  {Smee}, {Smith}, {Snedden}, {Stone}, {Stoughton}, {Strauss}, {Stubbs},
  {SubbaRao}, {Szalay}, {Szapudi}, {Szokoly}, {Thakar}, {Tremonti}, {Tucker},
  {Uomoto}, {Vanden Berk}, {Vogeley}, {Waddell}, {Wang}, {Watanabe},
  {Weinberg}, {Yanny}, {Yasuda}, \& {SDSS Collaboration}}]{sdss00}
{York}, D.~G., {Adelman}, J., {Anderson}, Jr., J.~E., {et~al.} 2000, \aj, 120,
  1579

\end{thebibliography}
%%%%%%%%%%%%%%%%%%%%%%%%%%%%%%%%%%%%%%%%%%%%%%%%%%

%%%%%%%%%%%%%%%%% APPENDICES %%%%%%%%%%%%%%%%%%%%%
% \appendix  

% \begin{figure*}
% \centering
% \includegraphics[scale=0.7]{images/Barbell_SI_er_map_c.pdf}
% \caption{The figure shows the spectral index error map 
% of Fig.\ \ref{fig:simap}.}
% \label{fig:siermap}
% \end{figure*}

%%%%%%%%%%%%%%%%%%%%%%%%%%%%%%%%%%%%%%%%%%%%%%%%%%%%%%%%%%%%%%%%%%%%%%%%%%%%%%%%

%%%%%%%%%%%%%%%%%%%%%%%%%%%%%%%%%%%%%%%%%%%%%%%%%
\end{document}